# Sleeping Beauties Cited in Patents: Is there also a Dormitory of Inventions?

Anthony F. J. van Raan
Centre for Science and Technology Studies
Leiden University
Wassenaarseweg 52
P.O. Box 9555
2300 RB Leiden, The Netherlands
vanraan@cwts.leidenuniv.nl

## *Abstract*

*A 'Sleeping Beauty in Science' is a publication that goes unnoticed ('sleeps') for a long time and then, almost suddenly, attracts a lot of attention ('is awakened by a prince'). In our foregoing study we found that roughly half of the Sleeping Beauties are application-oriented and thus are potential Sleeping Innovations. In this paper we investigate a new topic: Sleeping Beauties that are cited in patents. In this way we explore the existence of a dormitory of inventions. To our knowledge this is the first study of this kind. We investigate the time lag between publication of the Sleeping Beauty and the first citation by a patent. We find that patent citation may occur before or after the awakening and that the depth of the sleep, i.e., citation rate during the sleeping period, is no predictor for later scientific or technological impact of the Sleeping Beauty. A surprising finding is that Sleeping Beauties are significantly more cited in patents than 'normal' papers. Inventor-author self-citation occurs only in a small minority of the Sleeping Beauties that are cited in patents, but other types of inventor-author links occur more frequently. We develop a comprehensive approach in different steps to explore the cognitive environment of Sleeping Beauties cited in patents. First, we analyze whether they deal with new topics by measuring the time-dependent evolution in the entire scientific literature of the number of papers related to both the precisely defined topics as well as the broader research theme of the Sleeping Beauty during and after the sleeping time. Second, we focus on the awakening by analyzing the first group of papers that cites the Sleeping Beauty. Third, we create concept maps of the topic-related and the citing papers for a time period immediately following the awakening and for the most recent period. Finally, we make an extensive assessment of the cited and citing relations of the Sleeping Beauty. We find that tunable co-citation analysis is a powerful tool to discover the prince and other important application-oriented work directly related to the Sleeping Beauty, for instance papers written by authors who cite Sleeping Beauties in both the patents of which they are the inventors, as well as in their scientific papers.*

## 1. Introduction

A 'Sleeping Beauty in Science' is a publication that goes unnoticed ('sleeps') for a long time and then, almost suddenly, attracts a lot of attention ('is awakened by a prince'). We refer to our recent foregoing paper (van Raan 2015) for an overview of the literature on Sleeping Beauties (SBs), particularly the earlier pioneering work on 'delayed recognition' and studies on different aspects such as the occurrence of SBs; field-specific studies; examples of awaked application-oriented work; different patterns in citation histories of SBs; extreme historical cases; reasons for awakening; influences of SBs in performance assessments by citation analysis; SBs in the work of Nobel Prize winners; and identification of the 'princes'.

In our foregoing paper we discussed the results of an extensive analysis of Sleeping Beauties in physics, chemistry, and engineering & computer science (in this paper referred to as the three main fields) in order to find out the extent to which Sleeping Beauties are application-oriented and thus are potential Sleeping Innovations. We found that more than half of the SBs are application-oriented.

In this paper we take a further step by investigating whether the Sleeping Beauties in physics, chemistry, and engineering & computer science are also cited in patents. Thus we explore the existence of a dormitory of inventions. We are particularly interested in finding out how the time lag between publication of the SBs and their citation in patent is related to the sleeping period, and whether the number of patent-cited SBs is higher or lower than an average 'normal' paper.

The structure of this paper is as follows. We will discuss in Section 2 the data collection with a novel search algorithm, the analytical method to define Sleeping Beauties, the definition of the three main fields, the measuring procedure, the selection of specific set of SBs and the matching of the SBs with patent citation data in order to find SBs cited in patents. In Section 3 we analyze the basic properties of SBs that are cited in patents. Important topics are the time lag between publication and patent citation, the fields and countries of origin, the time-dependence of numbers and citation characteristics. Also we investigate inventor-author self-citation. Section 4 discusses the cognitive environment of the SBs cited in patents. Here we address the question whether SBs cited in patents deal with new, emerging topics. By taking the most highly cited SB as example, we analyze the awakening on the basis of the first group of citing papers. Concept maps as well as citation-based maps of the topic-related and the citing papers are created to further analyze characteristics of the cognitive environment, particularly to discover the prince and other important application-oriented work directly related to the SB and the patents in which it is cited. Moreover, further steps in the identification of inventor-author links are discussed.

## 2. Data, Method and Measuring Procedure

### 2.1 Publication data, variables and choice of specific set of Sleeping Beauties

In the foregoing paper (van Raan 2015) we discussed a fast and efficient Sleeping Beauty search algorithm written in SQL which can be applied to the CWTS enhanced Web of Science (WoS) database. With this algorithm we can tune the following four main variables: (1) *length of the sleep* in years after publication ($\boldsymbol{s}$); (2) *depth of sleep* in terms of a maximum citation rate during the sleeping period ($\boldsymbol{cs}_{max}$); (3) *awake* period in years after the sleeping period ($\boldsymbol{a}_{min}$ and $\boldsymbol{a}_{max}$); and (4) *awake intensity* in terms of a minimum citation rate during the awake period ($\boldsymbol{ca}_{min}$). We define $\boldsymbol{cs}_{max}$=0 as a coma, $\boldsymbol{cs}_{max}$= 0.5 as a very deep sleep, and $\boldsymbol{cs}_{max}$=1.0 as a deep sleep.

The algorithm allows selection of sets of WoS journal categories and thus restrict the search for Sleeping Beauties to one or more specific (main) fields of science, in this case physics, chemistry, and engineering & computer science. All variables can be tuned through any possible range of values, so that a continuum of Sleeping Beauties is found, ranging from 'mild' to 'extreme' ones. The total period in which the SBs and their citation data are searched for is 1980-2015, around 45,000,000 publications. The data analysis is carried out with the CWTS bibliometric database which is an improved and enriched version of the WoS database. Publication and citation data are available from 1980. The CWTS bibliometric database allows corrections for self-citations with high precision and provides a highly accurate unification of author names and institutions[1].

In the foregoing paper we investigated a set of physics SBs with (1) sleeping period length $\boldsymbol{s}$=10 years (publication years starting in 1980), (2) deep sleep, $\boldsymbol{cs}_{max}$ =1.0; (3) awake period of 10 years, $\boldsymbol{a}_{min}$= $\boldsymbol{a}_{max}$ = 10; and (4) awake intensity $\boldsymbol{ca}_{min}$ = 5.0. Thus, 1994 is the last year for publications having in total a twenty year time span (10 years sleep, awake period of 10 years) until 2013. We denote the SBs with these variables with [10, 1.0, 10, 5.0] as described in the foregoing paper. The number of these SBs for physics is 389. In a similar analysis for chemistry 265 SBs were identified, and for engineering & computer science 367.

In this study we combine the sets of the above physics, chemistry and engineering & computer science SBs in one combined set and focus on the most recent SBs within the defined time frame, namely those with publication years 1992-1994. For physics there are 122 of these 1992-1994 SBs, for chemistry 80 and for engineering & computer science 150. Although the publication of these SBs is more than 20 years ago, many of them started to get scientific and technological impact only recently.

## 2.2 Definition of the main fields

The main fields are defined as described in the foregoing paper (van Raan 2015). Physics, chemistry and engineering & computer science are main fields composed of (sub)fields as indicated in Table A1 (Appendix). The WoS journal-category codes given in this table are field identifiers used in our CWTS enhanced bibliometric WoS-based data system. The search algorithm selects the Sleeping Beauties in the fields defined by these identifiers. Notice that we use a broad definition of physics by including materials science and astronomy & astrophysics. Chemical engineering is included in the main field engineering & computer science.

## 2.3 Patent data and identification of SNPRs

Patents are documents with a legal status to describe and claim technological inventions in which, similar to scientific publications, references are given. These references concern mainly earlier patents ('prior art') in order to prove novelty in view of the existing technological developments and, generally to a lesser extent, to non-patent items, particularly scientific publications, the scientific non-patent references (SNPRs). References in scientific publications are the sole responsibility of the authors. The references in patents, however, can be given by both the inventors as well as by the patent examiners.

Clearly, these SNPRs represent a bridge between science and technology although they do not necessarily indicate the direct scientific basis of the invention described in the patent. Nevertheless, many studies (for an overview see for instance Callaert et al 2014) emphasize the importance of further research of the role of SNPRs in relation to the patented technological invention. In this study we focus on a particular phenomenon, namely the extent to which Sleeping Beauties show up as SNPRs (indicated as SB-SNPRs).

[1] CWTS, Data Infrastructure, http://www.cwts.nl/About-CWTS#Data_Infrastructure.

Patent publications were gathered by searching the EPO Worldwide Patent Statistical Database (PATSTAT), 2012 version. We group patent publications describing the same invention in 'patent families' to prevent double counting. In order to find out whether an SB is cited by patents, we matched all SBs on the basis of their WoS UT-codes with the citations given in patents. For more details we refer to Winnink and Tijssen (2014).

# 3. Basic Properties of Sleeping Beauties Cited in Patents

## 3.1 Time lag between publication and patent citation

In the set of 389 physics SBs we found 62 SNPRs, which is 16% of the total number of physics SBs. This first analysis is already a surprise. On average about 4% of all WoS-covered publications is an SNPR. Thus, the finding that 16% of our physics SBs is an SNPR means that these SBs are significantly more cited in patents than an average paper! The time lag between the publication year of the SB-SNPR and the first year of citation in a patent (***pcy***) ranges from 1 to 29 years, average 13.7 (sd=6.2). The most extreme case (***pcy***=29) is an SB from 1984 on the elastic properties of polymer composites which was cited in a patent not earlier than 2013. We call this year 'the most extreme year' (***Y***). Selecting the SBs with publication years 1992-1994, in total 122, 19 (again 16%) are identified as SNPRs. Here ***pcy*** ranges from 4 to 14.

In the set of 265 chemistry SBs, 62 SNPRs were found. This is 23%, which is even higher than the SNPR-percentage for physics. The ***pcy*** ranges from 1 to 29, average 12.4 (sd =5.6). The most extreme case is the same as the one in physics mentioned above because this SB is published in the journal Polymer Composites which is assigned to both physics and chemistry. In the subset of the 1992-1994 chemistry SBs, in total 80, also 19 are identified as SNPR, which is 24%. The ***pcy*** ranges from 1 to 19.

In the set of 367 Engineering and Computer Science we identified 108 SNPRs, which is 29%. This percentage is, as can be expected, higher than in physics and chemistry, but it is again surprisingly high. The ***pcy*** ranges from 1 to 27, average 11.8 (sd=5.6). Here the most extreme case concerns two SBs. One is from 1984 on the deformation of material at high temperature which was cited not earlier than 2011. The other is from 1985 on the generation of female sex hormones by plant-derived food which receives its first patent citation in 2012. In the 1992-1994 subset with 150 SBs, 30 (20%) are identified as SNPR, the ***pcy*** ranges from 2 to 18.

We measured the average ***pcy*** for the successive 3-years periods 1980-1982, 1983-1985, 1986-1988, 1989-1991, 1992-1994. The results are given in Table 1 and Fig 1. The data reveal a remarkable phenomenon: the average time lag between the publication year of an SB-SNPR and its first citation in a patent appears to decrease with about five years in a time period of 15 years. We have to be careful with this conclusion given the relatively large standard deviations.

| physics | N | pcy | sd | Y | | chemistry | N | pcy | sd | Y | | eng&comp | N | pcy | sd | Y |
|---|---|---|---|---|---|---|---|---|---|---|---|---|---|---|---|---|
| 1980-1982 | 10 | 17.2 | *5.2* | 2005 | | 1980-1982 | 15 | 14.7 | *5.6* | 2008 | | 1980-1982 | 13 | 12.4 | *8.0* | 2007 |
| 1983-1985 | 8 | 16.0 | *8.5* | 2013 | | 1983-1985 | 22 | 14.3 | *8.4* | 2013 | | 1983-1985 | 18 | 14.7 | *7.5* | 2012 |
| 1986-1988 | 11 | 13.0 | *6.9* | 2012 | | 1986-1988 | 10 | 13.1 | *5.8* | 2008 | | 1986-1988 | 17 | 13.6 | *6.7* | 2010 |
| 1989-1991 | 14 | 13.0 | *5.9* | 2013 | | 1989-1991 | 26 | 11.1 | *5.8* | 2012 | | 1989-1991 | 30 | 11.3 | *5.0* | 2011 |
| 1992-1994 | 19 | 11.9 | *3.9* | 2012 | | 1992-1994 | 19 | 10.1 | *6.3* | 2012 | | 1992-1994 | 30 | 9.1 | *5.7* | 2011 |

*Table 1. Average time lag with standard deviation between publication year and the first year of citation in a patent (**pcy**) for all SB-SNPRs in the given 3-years period (standard deviations are given in column **sd**). **N** is the number of SB-SNPRs and **Y** is the most extreme year in the given 3-years period.*

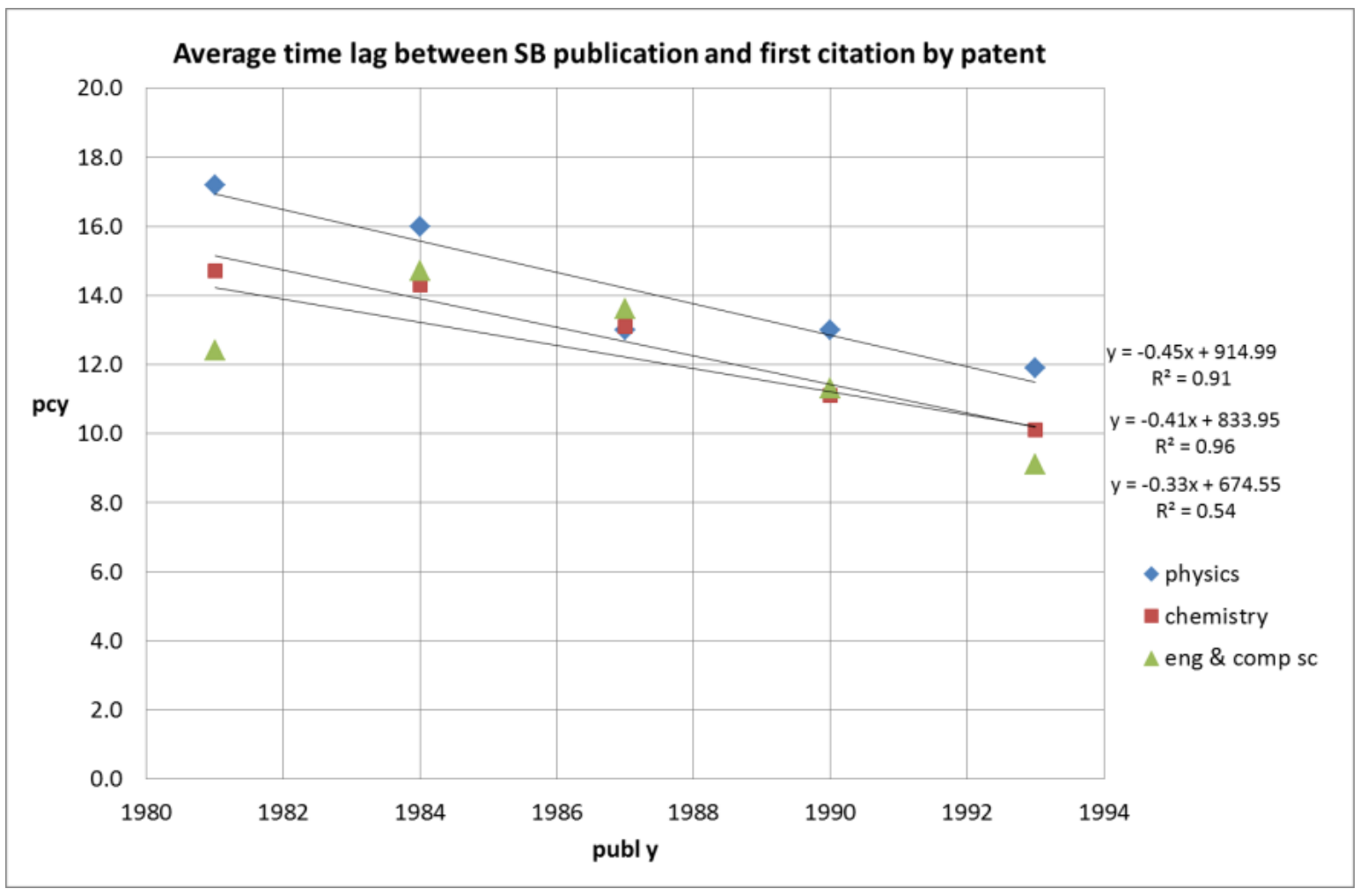

*Fig 1. Average time lag with standard deviation between publication year and the first year of citation in a patent (**pcy**) for all SB-SNPRs in the period 1980-1994 (values are given in the middle year of each 3-years period).*

Our current work on more recent, shorter sleeping SB-SNPRs appears to confirm the above observed trend. This would mean that SBs with technological importance, perhaps potential inventions, are 'discovered' increasingly earlier (van Raan 2016).

## 3.2 Fields and countries

As discussed earlier, our combined set of physics, chemistry and engineering & computer science SBs with publications years 1992, 1993 or 1994 covers 68 SB-SNPRs. Of these 68, 11 are in more than one of the main fields, so there are 57 unique SB-SNPRs. In Table 2 we present the distribution of these 57 SB-SNPRs over the different physics, chemistry and engineering & computer science fields (fields with 2 SNPRs or more).

For the fields with 5 or more SB-SNPRs we compared their share in the total number of 1992-1994 SBs (298) as well as in the total number of all publications 1992-1994 (1,018,821) in the three main fields. These shares are: Electrical & Electronic Engineering 11.7% of all SBs and 7.1% of all publications; Applied Physics 7.0% of all SBs and 6.8% of all publications; Optics 4.7% of all SBs and 3.1% of all publications; Energy Fuels 5.4% of all SBs and 1.7% of all publications; and Biochemistry & Molecular Biology 5.0% of all SBs and 15.6% of all publications.

We see that particularly Electrical & Electronic Engineering is overrepresented in SBs as well as in SB-SNPRs. Applied Physics, Optics, and Energy Fuels appear to be overrepresented in SB-SNPRs. Biochemistry & Molecular Biology is underrepresented in SBs as well as in SB-SNPRs. Given the low numbers, we however have to be careful with these conclusions. For the other fields in Table 2 the number of SB-SNPRs is too low for a meaningful conclusion about over- or underrepresentation.

| Fields | SB-SNPRs | % of SB-SNPRs |
|---|---|---|
| ENGINEERING ELECTRICAL ELECTRONIC | 10 | 17.5 |
| PHYSICS APPLIED | 6 | 10.5 |
| OPTICS | 6 | 10.5 |
| ENERGY FUELS | 5 | 8.8 |
| BIOCHEMISTRY MOLECULAR BIOLOGY | 5 | 8.8 |
| ENGINEERING BIOMEDICAL | 4 | 7.0 |
| CHEMISTRY MULTIDISCIPLINARY | 4 | 7.0 |
| BIOTECHNOLOGY APPLIED MICROBIOLOGY | 4 | 7.0 |
| ENGINEERING CIVIL | 3 | 5.3 |
| ENGINEERING CHEMICAL | 3 | 5.3 |
| CHEMISTRY MEDICINAL | 3 | 5.3 |
| REMOTE SENSING | 2 | 3.5 |
| PHARMACOLOGY PHARMACY | 2 | 3.5 |
| MATERIALS SCIENCE MULTIDISCIPLINARY | 2 | 3.5 |
| IMAGING SCIENCE PHOTOGRAPHIC TECHNOLOGY | 2 | 3.5 |
| GEOCHEMISTRY GEOPHYSICS | 2 | 3.5 |
| FOOD SCIENCE TECHNOLOGY | 2 | 3.5 |
| ENVIRONMENTAL SCIENCES | 2 | 3.5 |
| ENGINEERING MULTIDISCIPLINARY | 2 | 3.5 |
| CRYSTALLOGRAPHY | 2 | 3.5 |
| COMPUTER SCIENCE INTERDISCIPLINARY APPLICATIONS | 2 | 3.5 |
| CHEMISTRY PHYSICAL | 2 | 3.5 |
| CHEMISTRY APPLIED | 2 | 3.5 |
| CHEMISTRY ANALYTICAL | 2 | 3.5 |
| BIOPHYSICS | 2 | 3.5 |
| AUTOMATION CONTROL SYSTEMS | 2 | 3.5 |
| AGRICULTURAL ENGINEERING | 2 | 3.5 |

*Table 2. Distribution of the SB-SNPRs over the different physics, chemistry and engineering & computer science fields (fields with 2 SNPRs or more).*

| Countries | SB-SNPRs | % of total |
|---|---|---|
| US | 20 | 35.1 |
| JAPAN | 10 | 17.5 |
| GERMANY | 6 | 10.5 |
| UK | 4 | 7.0 |
| CANADA | 3 | 5.3 |
| AUSTRALIA | 3 | 5.3 |
| SWEDEN | 2 | 3.5 |
| SPAIN | 2 | 3.5 |
| NETHERLANDS | 2 | 3.5 |
| FRANCE | 2 | 3.5 |

*Table 3. Distribution of the SB-SNPRs over countries.*

The distribution of the 57 SB-SNPRs over countries is given in Table 3. More than half of the SB-SNPRs originates from the US and Japan. We compared for the countries with 5 or more SB-SNPRs their share in the total number of 1992-1994 SBs (298) as well as in the total number of all publications 1992-1994 (1,018,821) in the three main fields. These shares are: US 31.9% of all SBs and 34.2% of all publications; Japan 9.1% of all SBs and 9.6% of all publications; and Germany 7.7% of all SBs and 8.0% of all publications. Particularly Japan stands out in being overrepresented in the number of SB-SNPRs. But also here in most cases the low numbers do not allow hard conclusions.

## 3.3 Time-Dependence of Numbers and of Citation Characteristics

We divided all SBs in the entire period 1980-1994 into two categories: the SBs that are not cited in a patent (SB-nonSNPR), and those that are cited in a patent (SB-SNPR). For both categories we measured for the successive 3-years periods 1980-1982, 1983-1985, 1986-1988, 1989-1991, 1992-1994 the numbers, the average number of citations per year during the 10 years sleeping period (***cs***), and the average number per year during the ten years awakenings after the sleeping period (***ca***).

First the time-dependence of the numbers, see Fig 2. The main fields physics, chemistry and engineering & computer science are combined. Given the low numbers we can expect that the significance of our findings is not high, but nevertheless we observe a small exponential increase (exponents are given in the figure) of both the number of SB-nonSNPRs (with an average exponent of 0.09) and SB-SNPRs whereby the increase of the SB-SNPRs seems to be smaller (with an average exponent of 0.05). This could imply a decreasing probability that a Sleeping Beauty is cited by a patent because the technological relevance of a publication is recognized before it can become a Sleeping Beauty.

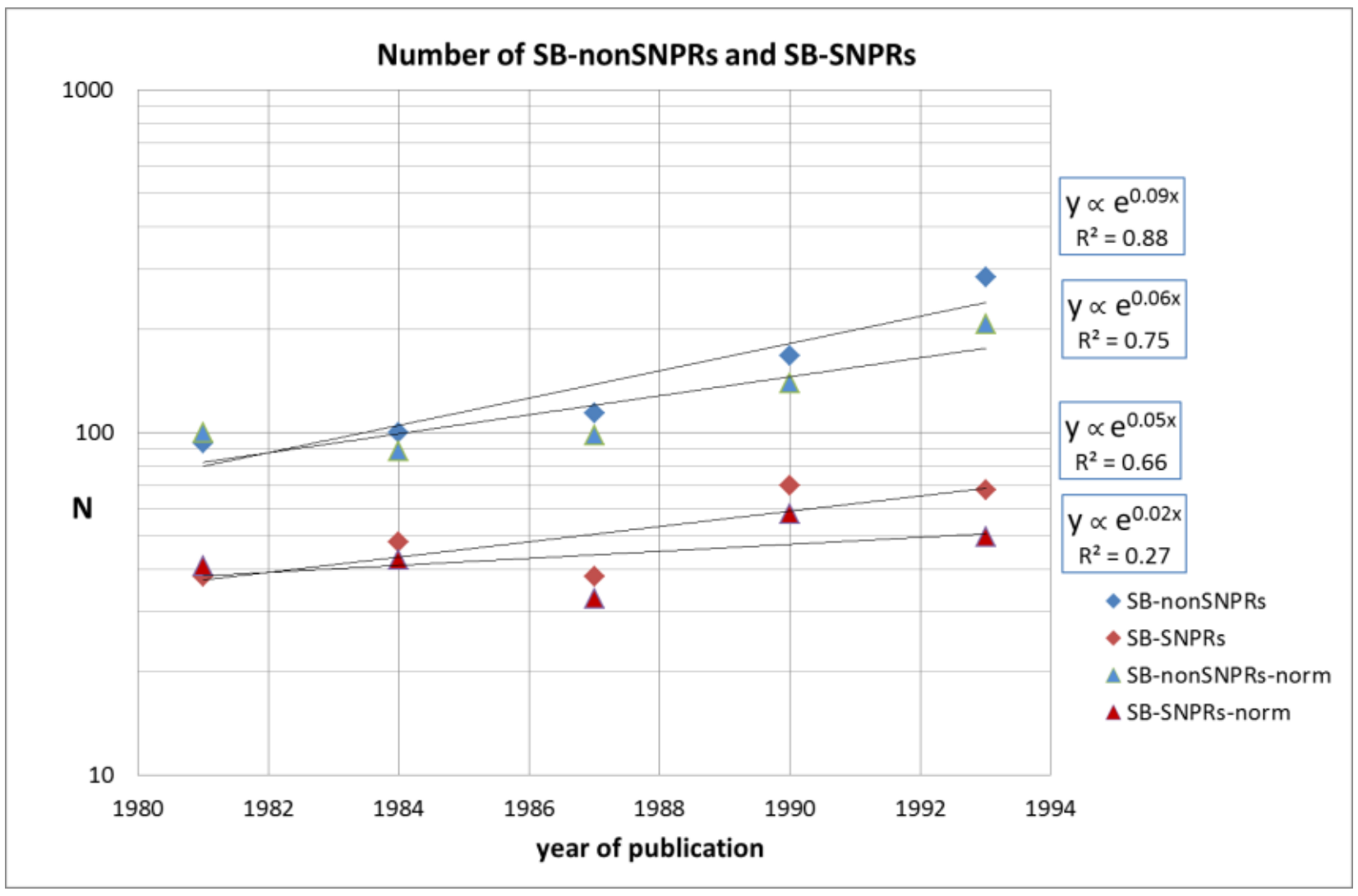

*Fig 2. Numbers of SB-nonSNPRs and of SB-SNPRs. We indicate the 3-year periods with the middle years, e.g., the numbers given for 1981 concern the period 1980-1982. Diamonds represent absolute values, triangles normalized values (blue: SB-nonSNPRs, red: SB-SNPRs).*

This finding is in line with our conclusions based on the data discussed in Section 3.1. This trend is even more visible in the normalized values (index SB-nonSNPRs 1980-1982

= 100): for the SB-nonSNPRs we see still an increase but less than in the case of absolute values, and for the SB-SNPRs there is hardly a significant increase.

In Table 4 we also present the numbers but now for each main field, together with the average number of citations per year during the 10 years sleeping period (***cs***), and the average number of citations per year during the ten years awakenings after the sleeping period (***ca***). For both the SB-nonSNPR as well as for the SB-SNPR we find a weak correlation between the two indicators which is, surprisingly, negative.

| **SB-nonSNPR** | ***N*** | ***cs*** | | | ***ca*** | |
|---|---|---|---|---|---|---|
| physics | | av | *sd* | | av | *sd* |
| 1980-1982 | 45 | 0.70 | 0.29 | | 6.63 | 1.30 |
| 1983-1985 | 47 | 0.68 | 0.27 | | 8.77 | 7.94 |
| 1986-1988 | 59 | 0.62 | 0.28 | | 7.40 | 2.82 |
| 1989-1991 | 73 | 0.69 | 0.24 | | 7.39 | 2.90 |
| 1992-1994 | 103 | 0.65 | 0.26 | | 7.45 | 3.78 |
| | | | | | | |
| chemistry | | av | *sd* | | av | *sd* |
| 1980-1982 | 25 | 0.71 | 0.26 | | 6.06 | 1.14 |
| 1983-1985 | 23 | 0.78 | 0.21 | | 6.45 | 2.10 |
| 1986-1988 | 21 | 0.74 | 0.26 | | 6.13 | 0.26 |
| 1989-1991 | 43 | 0.73 | 0.25 | | 6.89 | 3.83 |
| 1992-1994 | 61 | 0.70 | 0.26 | | 6.76 | 2.15 |
| | | | | | | |
| eng & c sc | | av | *sd* | | av | *sd* |
| 1980-1982 | 23 | 0.79 | 0.22 | | 6.34 | 1.07 |
| 1983-1985 | 30 | 0.78 | 0.24 | | 6.31 | 1.59 |
| 1986-1988 | 34 | 0.71 | 0.25 | | 6.57 | 1.69 |
| 1989-1991 | 52 | 0.68 | 0.26 | | 6.34 | 1.63 |
| 1992-1994 | 120 | 0.72 | 0.24 | | 6.69 | 1.92 |

| **SB-SNPR** | ***N*** | ***cs*** | | | ***ca*** | |
|---|---|---|---|---|---|---|
| physics | | av | *sd* | | av | *sd* |
| 1980-1982 | 10 | 0.59 | 0.38 | | 8.40 | 3.96 |
| 1983-1985 | 8 | 0.74 | 0.20 | | 6.63 | 1.65 |
| 1986-1988 | 11 | 0.72 | 0.26 | | 6.31 | 1.20 |
| 1989-1991 | 14 | 0.71 | 0.19 | | 7.29 | 2.60 |
| 1992-1994 | 19 | 0.78 | 0.26 | | 7.92 | 2.74 |
| | | | | | | |
| chemistry | | av | *sd* | | av | *sd* |
| 1980-1982 | 15 | 0.67 | 0.34 | | 7.12 | 1.92 |
| 1983-1985 | 22 | 0.72 | 0.22 | | 6.86 | 1.46 |
| 1986-1988 | 10 | 0.58 | 0.22 | | 7.88 | 2.51 |
| 1989-1991 | 26 | 0.70 | 0.24 | | 6.75 | 2.86 |
| 1992-1994 | 19 | 0.69 | 0.24 | | 7.36 | 2.66 |
| | | | | | | |
| eng & c sc | | av | *sd* | | av | *sd* |
| 1980-1982 | 13 | 0.74 | 0.29 | | 6.73 | 1.51 |
| 1983-1985 | 18 | 0.76 | 0.23 | | 6.22 | 0.93 |
| 1986-1988 | 17 | 0.75 | 0.25 | | 6.39 | 1.08 |
| 1989-1991 | 30 | 0.74 | 0.22 | | 6.08 | 1.15 |
| 1992-1994 | 30 | 0.75 | 0.24 | | 8.01 | 4.32 |

*Table 4. Number (**N**), average citation rate during the 10 years sleeping period (**cs**), and average citation rate during the 10 years awakening period (**ca**) for SB-nonSNPRs (left part) and for SB-SNPRs (right part). Standard deviations are given in the columns sd.*

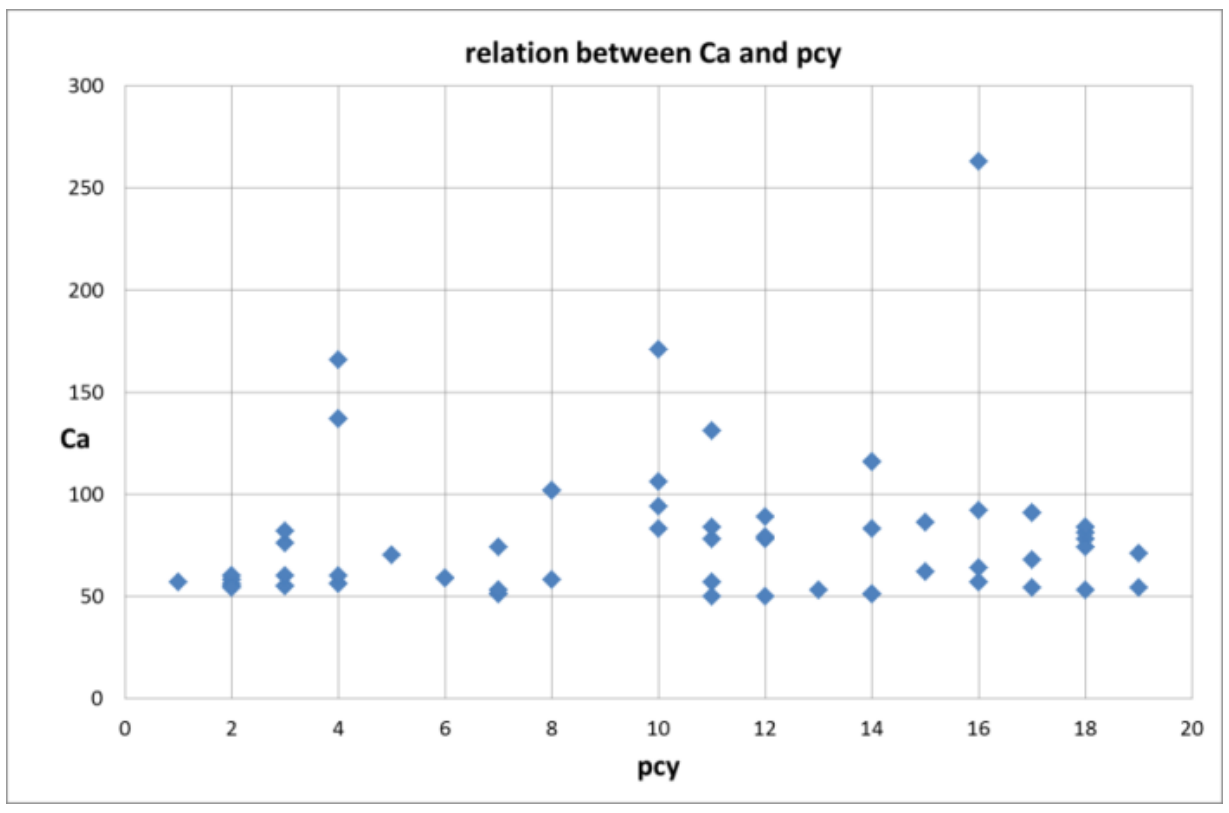

*Fig 3. Relation of the time lag between publication year and the first year of citation in a patent (**pcy**) and the total number of citations received in the 10-years awakening period (**Ca**).*

Is there a relation of the time lag between publication year and the first year of citation in a patent (***pcy***) and the total number of citations received in the 10-years awakening period (***Ca***)? Fig 3 shows that there is no relation. Also the highly cited SBs have ***pcy*** values in the whole range of time lags. Thus we conclude that the citation rate during the sleeping period is no predictor for later scientific or technological impact of the Sleeping Beauty.

## 3.4 Inventor-author self-citations

An important bridge between science and technology is built on the work of scientists who are both inventors of patented technology as well as authors of publications. The number of studies on inventor-author relations is quite limited. One of the few early studies is our work on inventor-author relations in the field of the application of lasers in medicine (Noyons et al. 1994).

For large scale studies of inventor-author relations the lack of unification of names and accurate person identification in publication- as well as patent databases poses a major problem. Therefore, most studies are still on a smaller scale. To our knowledge there are only two large-scale studies. The CWTS–Fraunhofer study on the development of nanoscience and nanotechnology in the EU countries (Noyons et al. 2003) is a very comprehensive inventor–author study. In this study over 15,000 inventor–authors combinations were identified with help of several text-analysis techniques. Boyack and Klavans (2008) investigated science–technology interaction on a large scale by identifying and validating a set of nearly 20,000 inventor–authors through matching of rare names obtained from paper and patent data. With rare names the probability to identify a single person is considerably higher than in the case of common names. But, of course, not all important inventor-authors have rare names and therefore this method is not generally applicable.

In this study we focus our inventor-author analysis on the 57 SB-SNPRs in the period 1992-1994. More particularly, we investigate the extent to which SB-SNPRs are cited in patents of which at least one of the inventors is also an author of the cited SB-SNPR. We call this inventor-author self-citation. We operate on a small scale and complicated matching techniques are not necessary. Simple semi-automatic comparison of the inventor names and the author names with manual checks are sufficient. We found that these 57 SB-SNPRs are cited in 143 patents families which cover in total 191 patents and 315 inventors. As an example of our data we show in Table 5 the two patent families with each three patents in which one of the most highly cited SB-SNPRs, Bonsignore et al (1993) (abbreviated BLSC SB-SNPR, further discussed in Section 4.1) is cited. In the table we find the WoS UT code of the BLSC SB-SNPR, the registration number of the patent families and of the patents (formally mentioned 'patent publications'); the title of the patents (we see that the same invention is patented in different patent offices); the names of the inventors; the IPC (International Patent Office) codes which indicate the relevant fields of technology, for instance A61K31/37 concerns organic compounds for medical purposes, particularly coumarins (coumarins are benzopyran-related compunds); and finally the application year.

In this case we see that there is no inventor-author combination, the names of the inventors are different from the SB-SNPR paper authors: L. Bonsignore, G. Loy, D. Secci, and A. Calignano. We see the similarity between the topic discussed in this SB-SNPR with title 'Synthesis and pharmacological activity of 2-Oxo-(2h) 1-benzopyran-3-carboxamide derivatives', and the titles of the patents.

| UT code | Patent family | Patent publ | Patent title | Inventors | IPC codes | Appl year |
|---|---|---|---|---|---|---|
| A1993LM31100009 | 9507531 | FR19970006814<br>WO1998FR01087<br>US20000445177 | Coumarin derivatives, methods of preparation and application as medicines | Delarge, Jacques<br>Doucet, Caroline<br>Boggetto, Nicole<br>Pirotte, Bernard<br>Pochet, Lionel<br>Reboud Ravaux, Michele | A61K31/37, A61K31/4433, A61K31/4436, A61K31/453, A61K31/4709, A61K31/4725, A61P1/18, A61P7/02, A61P9/10, A61P11/00, A61P13/12, A61P17/00, A61P19/02, A61P29/00, A61P31/00, A61P35/00, A61P43/00, C07D311/14, C07D335/06, C07D405/12, C07D409/12 | 2000 |
| A1993LM31100009 | 32400079 | EP20030776786<br>US20050537711<br>EP20100190730 | Novel 2H-chromen-2-one-3-carboxamides for medical uses | Chen, Xiaoguang<br>Cheng, Guifang<br>Li, Hongyan<br>Li, Lanmin<br>Li, Yan<br>Xie, Longfei<br>Xu, Shiping<br>Xu, Song | A61P3/10, A61P9/10, A61P9/12, A61P13/12, A61P35/00, C07D311/16, C07D405/12, C07D413/04, C07D413/14, C07D417/12, A61K31/366, A61K31/416, A61K31/4245, A61K31/4439, A61K31/513, A61K31/585, C07D311/00, C07D311/02, C07D405/02, A61K31/37, C07D311/20 | 2005 |

*Table 5. Example of the data of patents in which the BLSC SB-SNPR is cited.*

By matching the 315 names (last name and initials) of the inventors with the 159 names (last name and initials) of the authors of the 57 SB-SNPRS we find that 4 authors are also inventors. Two authors are co-authors as well as co-inventors. This means that for 3 of the 57 SB-SNPRs (about 5%) an inventor-author relation is established.

In Table 6 we give en example of such an inventor-author relation: the 1994 SB-SNPR of A. Moreira and Y.H. Huang on the processing of airborne high-resolution radar signals (title 'Airborne SAR processing of highly squinted data using a chirp scaling approach with integrated motion compensation', Moreira and Huang 1994). There are two patent families, one with two patents and one with one patent in which the Moreira and Huang SB-SNPR is cited. We see that for this latter patent the author Moreira is also one of the inventors. The other two patents have an inventor who is not an author of the cited SB-SNPR.

| UT code | Patent family | Patent publ | Patent title | Inventors | IPC codes | Appl year |
|---|---|---|---|---|---|---|
| A1994PF41500008 | 4169219 | WO2002CA00886<br>US20030730189 | Imaging system utilizing spatial image oscillation | Zador, Andrew | G06T1/00, H04N5/217, H04N5/349, H04N5/357, G03B13/00, H04N5/232 | 2003 |
| A1994PF41500008 | 7788072 | US19970816044 | Method for azimuth scaling of SAR data and highly accurate processor for two-dimensional processing of scanSAR data | Mittermayer, Josef<br>Moreira, Alberto | G01S13/90 | 1997 |

*Table 6. Example of the data of patents in which an author of the cited SB-SNPR is also an inventor.*

Also in this case we notice the similarity between the topic discussed in this SB-SNPR and the titles of the patents. IPC-code G06T1/00 covers image processing techniques in general, G01S13/90 concerns systems using the reflection or re-radiation of radio waves, e.g. radar systems, with synthetic aperture techniques.

We conclude that only for a small minority (5%) of the Sleeping Beauties that are cited in patents the authors are also inventors of the technology described in the citing patent. We remind that this type of inventor-author link can be regarded as an inventor-author self-citation. It is very well possible that authors of SB-SNPRs are inventors of other patents than the patents that cite the author's SB-SNPR. We come back to this issue with examples in Section 4.5 and in forthcoming work we will investigate in more detail to what extent authors of Sleeping Beauties are inventors of technological developments described in later patents.

# 4. Cognitive Environment of Sleeping Beauties Cited in Patents

## 4.1 Do Sleeping Beauties Cited in Patents Deal with New Topics?

A central question is whether the SB-SNPRs were dealing with new topics that were hardly or not the subject of other publications. Therefore we analyzed for the 5 most cited (after awakening) SB-SNPRs the time-dependent evolution of their topics in the entire scientific literature. We first identified the topics of the SB-SNPRs by using meaningful concepts from the title and abstracts. Next, we uploaded a search string with these concepts (which defines the topic of an SB-SNPR) into the Web of Science for the period 1991[2]-2015 and determined the annual number of publications dealing with the same topic as the SB-SNPR. This time evolution is then compared with the citation history of the SB-SNPR.

We illustrate our approach first with the most cited SB-SNPR, the paper of Tassiulas and Ephremides (TE) published in 1992 on stability properties of constrained queuing-systems and scheduling policies for maximum throughput in multihop radio-networks (Tassiulas and Ephremides 1992)[3]. Meanwhile this paper is cited more than 500 times (April 7, 2016: 533 citations[4]), which is quite exceptionally, certainly for a Sleeping Beauty. We classified this paper in the main field Engineering & Computer Science and it was published in the journal IEEE Transactions on Automatic Control.

For identification of other papers directly related to the TE SB-SNPR, we used the search string [queu* AND (multihop* OR multi hop OR multi-hop*) AND network*]. This search resulted in the identification of 266 papers in the period between 1991 and 2015. The results are presented in Fig 4 together with the number of citations to the TE SB-SNPR. We see that during the sleeping period of the TE SB-SNPR, the scientific interest in the topic concerned was very low. The figure shows that 2003 is the awakening year. From that year the scientific interest in the topic rapidly increased, later on followed by technological interest demonstrated by the first patent citation. This first citation in a patent occurred 5 years after the awakening, which is 16 years after the publication of the TE SB-SNPR. Fig 4 shows that although the patent citation did not trigger the awakening, it may have reinforced also the scientific interest in the TE paper given the somewhat steeper increase in citations.

---

[2] Before 1991 only titles of papers and not abstracts are available for topic searches in the WoS. Therefore, we take 1991 as the starting year for our analysis.
[3] In a multi-hop wireless network, communication between the begin node and the end node of a network runs through a number of intermediate nodes that relay information from one node to another.
[4] Citations by research papers included in the WoS, Science Citation Indexes expanded, thus not citations by conference papers and books.

A remarkable observation is that the number of papers citing the TE SB-SNPR after awakening is considerably larger than the topic-related papers. This shows that the TE SB-SNPR also has an impact on research outside the work directly related to the topic. We will see later on that this TE SB-SNPR is now considered as one of the founding works on stochastic network optimization. To answer the question posed in the title of this section: this SB-SNPR clearly dealt with a new topic.

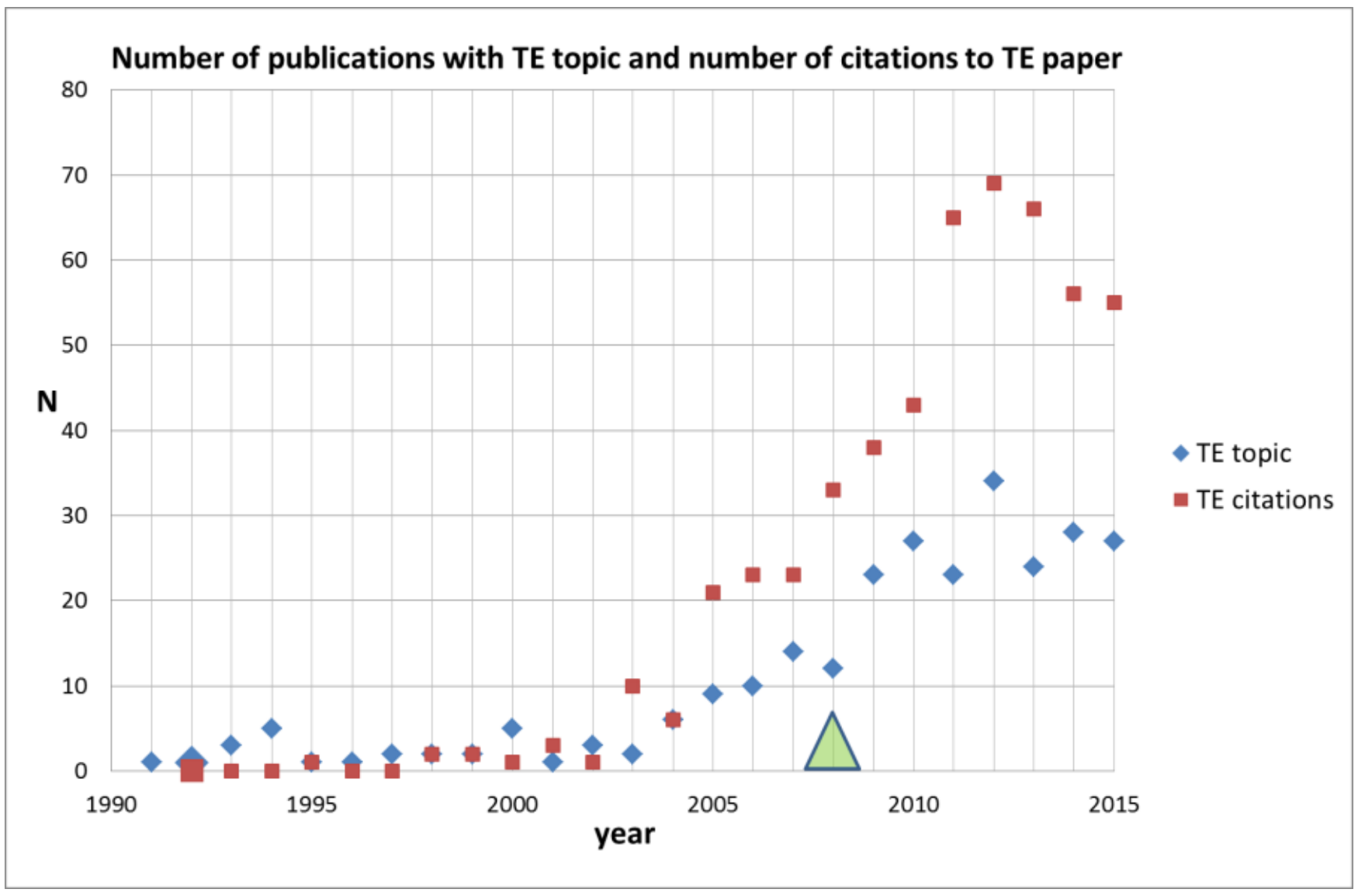

*Fig 4. Tassiulas & Ephremides SB-SNPR (deep sleep **cs**=0.9). Red squares indicate number of citations, blue diamonds indicate number of topic-related papers. The large blue diamond and large red square indicate the publication year (1992) of the TE paper. The green triangle indicates the year of first citation in a patent (2008).*

The next SB-SNPR of the top-5 is Birks and Li (BL) published in 1992 on the shape of fiber tapers, an important topic in research and technology on the propagation of light in fibers (Birks and Li 1992). This paper is classified in the main fields Physics as well as in Engineering & Computer Science and it was published in the Journal of Lightwave Technology. Meanwhile it has been cited 277 times (April 7, 2016).

For identification of other papers directly related to the BL SB-SNPR, we used the search string [shape* AND fiber* AND taper*]. This search resulted in the identification of 380 papers in the period between 1991 and 2015. The results are presented in Fig 5 together with the number of citations to the BL SB-SNPR. We see that during the sleeping period the scientific interest in the topics concerned was already present, but it increased rapidly with the awakening of the BL SB-SNPR in 2003.

Although in this case the SB-SNPR did not deal with an entirely new topic, but it presented new technological applications: the first citation in a patent occurred already 4 years after publication. Thus, the 'technological awakening' is earlier than the scientific awakening, which is 7 years later. Moreover, our data suggest that this technological awakening triggered the awakening and subsequent growth of scientific interest in the topic.

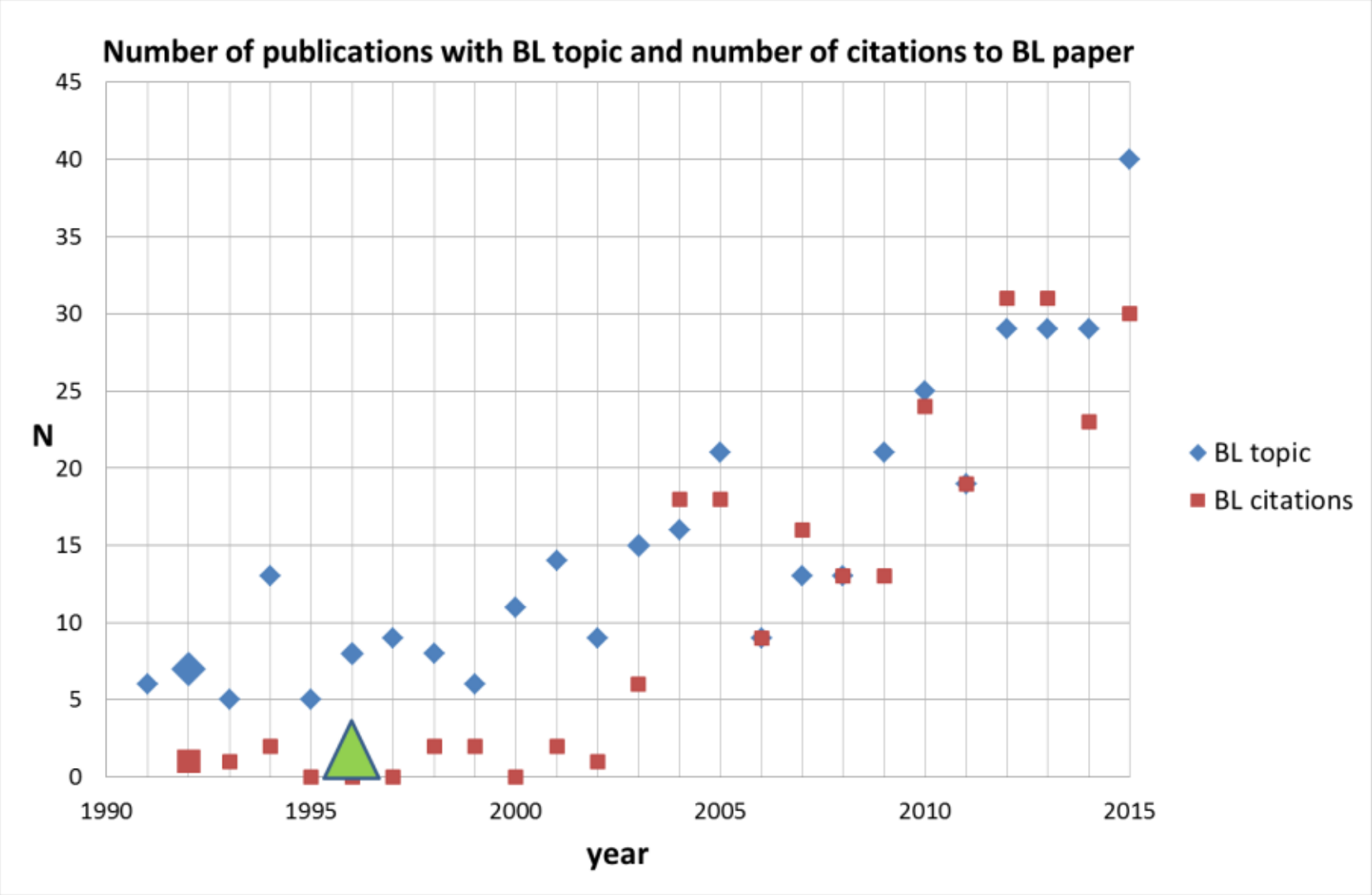

*Fig 5. Birks and Li SB-SNPR (deep sleep **cs**=1.0). Explanation see Fig 4.*

To avoid overloading this paper we do not discuss the details of the awakening of the BL and the three following SB-SNPRs. We take the TE SB-SNPR as a general example of a detailed analysis of the awakening in Section 4.

The third SB-SNPR of the top-5 is Chandorkar, Divan and Adapa (CDA) published in 1993 on the control of parallel connected inverters in standalone AC supply-systems (Chandorkar et al 1993). Meanwhile it has been cited 179 times (April 7, 2016). This paper is classified in the main field Engineering & Computer Science and it was published in the IEEE Transactions on Industry Applications.

For identification of other papers directly related to the CDA SB-SNPR, we used the search string [control AND parallel AND inverter* AND AC]. This search resulted in the identification of 234 papers in the period between 1991 and 2015. The results are presented in Fig 6 together with the number of citations to the CDA SB-SNPR. We see that, quite similar to the foregoing case, some scientific interest in the topics concerned was already present during the sleeping period. Although formally the CDA paper satisfies the SB-definition [10, 1.0, 10, 5.0], it took a considerably longer time to become 'awake'. As can be seen in Fig 6, not earlier than in 2010 the number of citations increased substantially. The first patent citation was in 2005, 12 years after the publication of the CDA SB-SNPR. Our data suggest that this first patent citation has triggered the scientific interest for the topic. As discussed above, it then took another few years before the interest in the CDA SB-SNPR in terms of citations started to increase.

Also here the topic of the SB-SNPR was not entirely new in the time this SB-SNPR was published. In contrast to the two foregoing cases, here our data suggest that the first patent citation triggered the interest for the topic and subsequently the interest for the CDA paper.

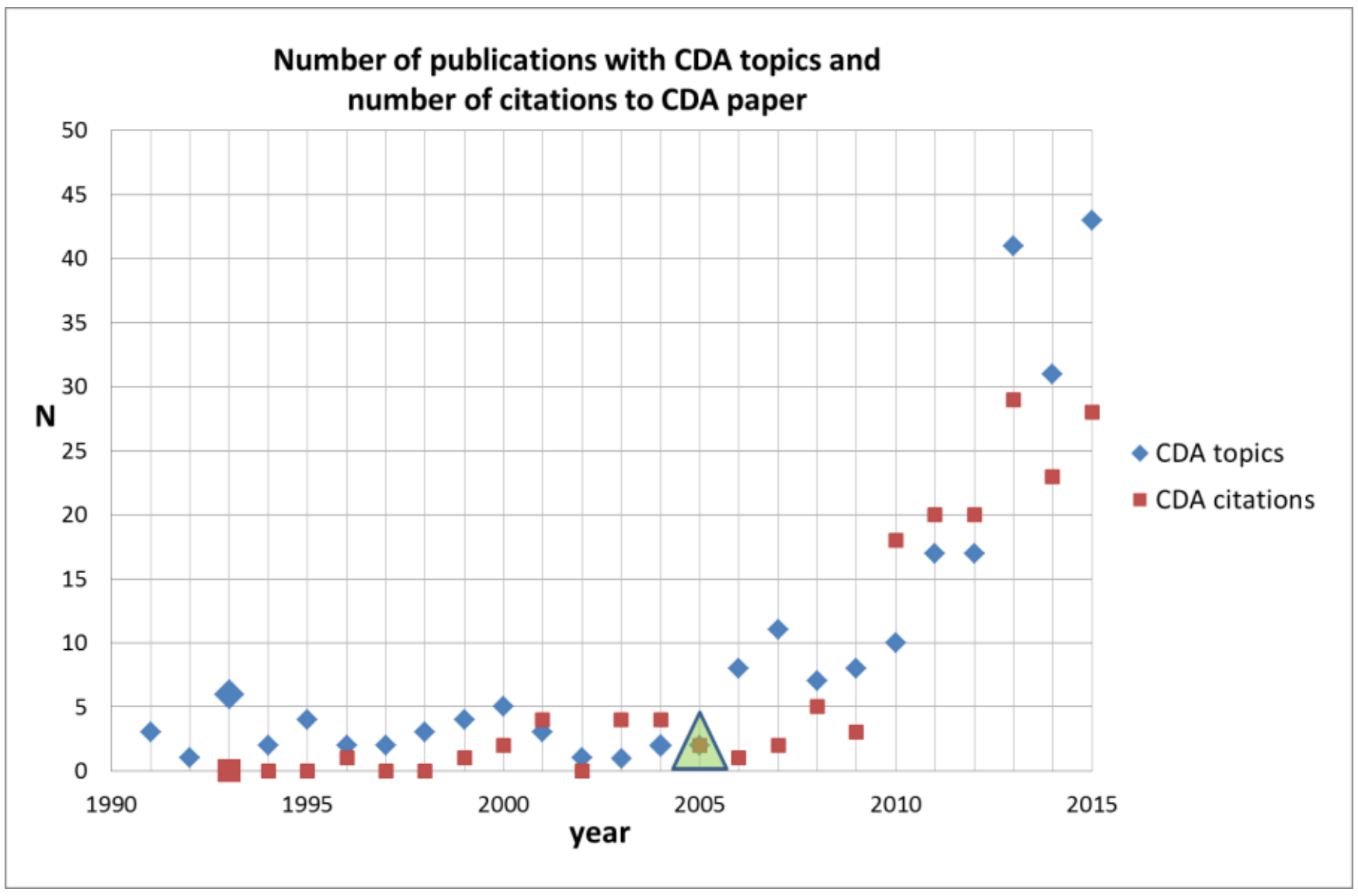

*Fig 6. Chandorkar et al SB-SNPR (deep sleep **cs**=0.8). Explanation see Fig 4.*

The fourth SB-SNPR of the top-5 is Bonsignore, Loy, Secci, and Calignano (BLSC) published in 1993 on the synthesis and pharmacological activity of benzopyran-carboxamide derivatives which play an important role in anticoagulant and diuretic medicines (Bonsignore et al 1993). Benzopyran derivates are also potentially useful ant-inflammatory and anti-cancer agents. This paper is classified in the main field Chemistry and it was published in the European Journal of Medicinal Chemistry. Meanwhile it has been cited 340 times (April 13, 2016).

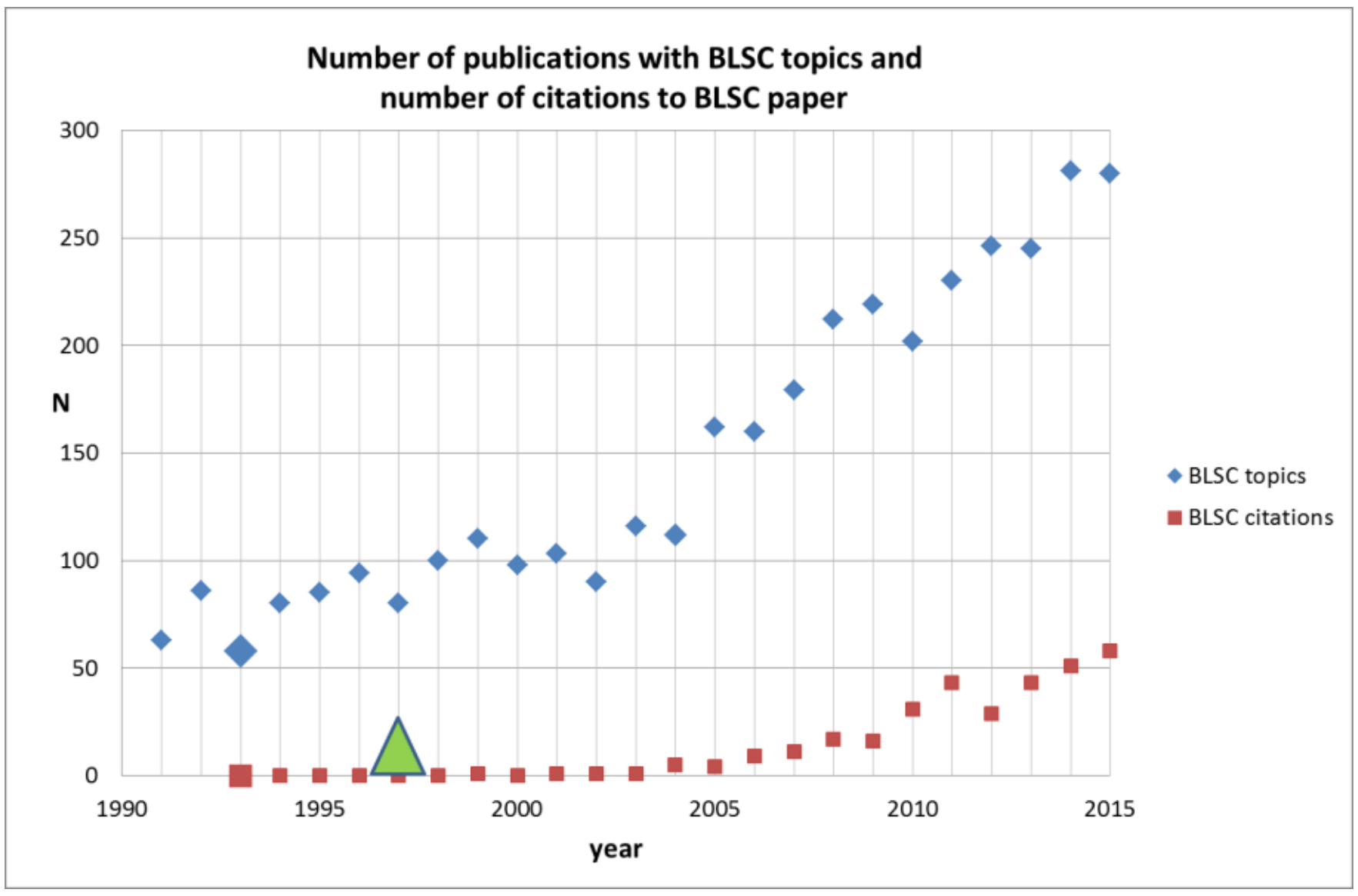

*Fig 7. Bonsignore et al SB-SNPR (very deep sleep **cs**=0.3). Explanation see Fig 4.*

For identification of other papers directly related to the BLSC SB-SNPR, we used the search string [(benzopyran* OR carboxamid*) AND synthe* AND derivat*]. This search resulted in the identification of 3,691 papers in the period between 1991 and 2015. The results are presented in Fig 7 together with the number of citations to the BLSC SB-SNPR. We see a remarkable phenomenon. The scientific interest in the topic increased around the BLSC publication year, but the BLSC paper remained practically uncited. However, the first patent citation –already 4 years after the BLSC publication- may have caused a further increase in the scientific interest as visible in Fig 7, whereas the BLSC paper is still hardly cited. Clearly, the first patent citation did not trigger the awakening of the BLSC paper. The figure also shows that the awakening of the BLSC SB-SNPR is not a sudden event: from 2004 the numbers of citations start to increase gradually. We also observe that together with this slow awakening the scientific interest in the topic increased considerably.

The last SB-SNPR of the top-5 is Li and Ahmadi (LA) published in 1992 on the dispersion and deposition of spherical particles from point sources in a turbulent channel flow (Li and Ahmadi 1992). This paper is classified in the main fields Physics as well as in Engineering & Computer Science and it was published in the journal Aerosol Science and Technology. Up till now this paper has been cited 262 times.

For identification of other papers directly related to the LA SB-SNPR, we used the search string [(dispersion OR deposition) AND *particle* AND source* AND turbulen* AND flow*]. This search resulted in the identification of 384 papers in the period between 1991 and 2015. The results are presented in Fig 8 together with the number of citations to the LA SB-SNPR. We see that there existed already scientific interest in the topics during the sleeping period. This interest seems to increase around the awakening. The figure shows that also in this case the awakening develops gradually. The first citation in a patent occurred 14 years after publication, and 4 years after the awakening of the LA SB-SNPR. Around the same time the scientific interest in the topic stabilized whereas the number of citations to the LA paper continued to increase. This indicates that the LA paper has become an important, leading paper in its research theme as a whole.

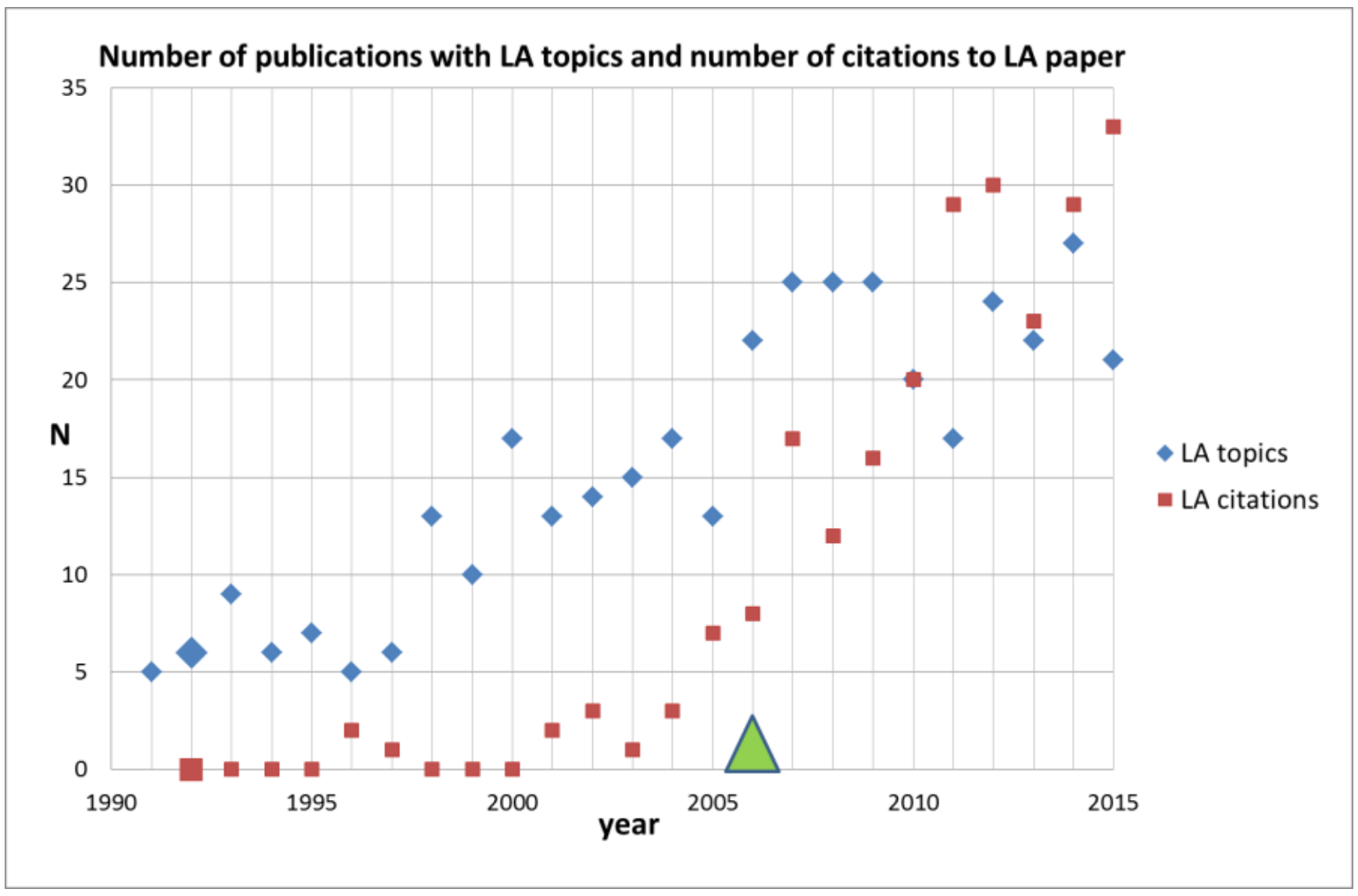

*Fig 8. Li and Ahmadi SB-SNPR (very deep sleep, **cs**=0.5). Explanation see Fig 4.*

On the basis of the above observations, we conclude that Sleeping Beauties that are cited in patents may deal with an entirely new topic but this is not a general rule. Rather they present within an existing research topic or research theme new approaches which pave the way to new applications and stimulate further research and development.

## 4.2 References and Early Citing Papers

Does the 'cognitive environment', operationalized in the citing and cited relations of SBs that are cited in patents, reveal more about the characteristics of SB-SNPRs, particularly details of the awakening? In order to investigate this, we present a general analytical method which we apply as an example to the most cited SB-SNPR, the TE paper.

As in our foregoing paper (van Raan 2015) we applied the CWTS bibliometric analytical tool CitNetExplorer (van Eck and Waltman 2014) to investigate (1) the papers cited by the SB-SNPR (i.e., its references) and (2) in order to focus on the awakening, its early (first 25) citing papers.

The papers cited by the TE SB-SNPR are located in the upper part of Fig. 9. Analysis of these TE references shows that the oldest are two highly cited books and a conference paper. The two books are on the theoretical foundations of the applications described in the TE SB-SNPR and in the patents citing TE: Kemeny et al (1976) on Markov chains, and Papadimitriou and Steglitz (1982) on combinatorial optimization. The conference paper by Silvester (1982) is on scheduling in multi-hop broadcast networks and directly related to the topic of the TE SB-SNPR.

A network map of the early, particularly the first 25, citing papers is shown in the middle part of Fig. 9. The CitNetExplorer also reveals the citation relations between these early citing papers. Evidently, the TE SB-SNPRs is the commonly cited paper. As discussed earlier, the TE paper is a SB with a 'deep sleep' (***cs***=0.9) in the 10 years after publications. This means that, without the two self-citations there are 9 citing publications in the period 1992-2001 of which 5 are from one and the same first author, S. C. Kam. These papers form a separate cluster on problems with network access and in particular on bandwidth guarantee. The Sarkar paper from 2002 is also a self-citation, Sarkar and Tassiulas 2002).

Clearly, from 2003 the awakening is a fact. But who is the prince, i.e., which paper triggered the awakening? Because of the self-citations, 'self-awakening' may also play a role. Often, and also in this case, there are more candidates. We think that particularly citing papers that are highly cited papers themselves (number of citations until now above 100) are the most important prince-candidates. We indicated these highly cited citing papers with a square. The first highly cited citing paper is McKeown (first author) in 1999 (McKeown et al 1999) which can be regarded as an 'early passing prince' without really triggering the awakening. Thus, more appropriate candidates are the 2003 papers of Neely (Neely et al 2003) and of Andrews (Andrews and Vojnovic 2003).

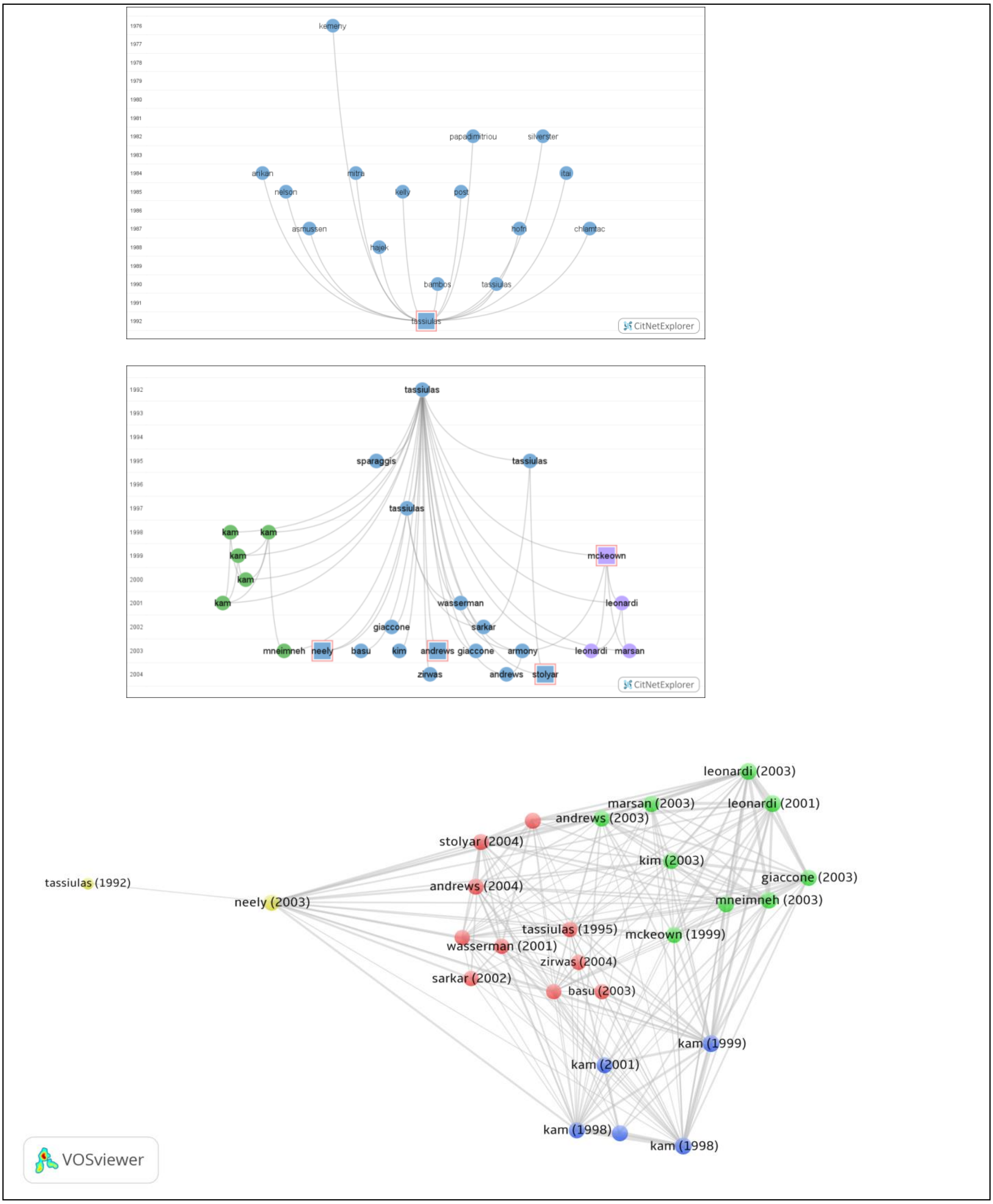

*Fig 9. Upper part: the references of the TE SB-SNPR; middle part: the 25 oldest papers citing the TE SB-SNPR; lower part: bibliographic coupling of these 25 oldest citing papers. The red circle between the Wasserman (Wasserman and Olsen 2001) and the Neely (Neely et al 2003) papers represents Tassiulas 1997.*

In the lower part of Fig. 9 we show the bibliographic coupling network of these first 25 citing papers combined with the TE SB-SNPR. We immediately see that the TE SB-SNPR (indicated as Tassiulas 1992 on the extreme left hand side of the map) has an 'isolated' position. This means that it has its own specific references which have a relatively low similarity with the references of the other papers. The exception is the Neely et al (2003) paper which forms a bridge between the TE SB-SNPR and all other early citing papers, particularly the red cluster which covers three other papers by Tassiulas (Tassiulas 1995, Tassiulas 1997, Sarkar and Tassiulas 2002) which self-cite the TE SB-SNPR. This confirms our opinion, as discussed above, that the Neely et al (2003) paper is a strong candidate for the role of the prince.

## 4.3 Broadening the research theme

The different clusters in Fig 9 suggest that the specific delineation of the 'cognitive environment' of the TE SB-SNPR applied in the foregoing section with the search string [queu* AND (multihop* OR multi hop* OR multi-hop*) AND network*] (in short: TE topic) can be broadened from a topic-related to a theme-related cognitive environment by removing 'queu*'. Indeed, the search string [(multihop* OR multi-hop*) AND (network*)] (in short: multihop) yields 4,767 publications between 1991 and 2015, almost a factor 20 more than the specific search string.

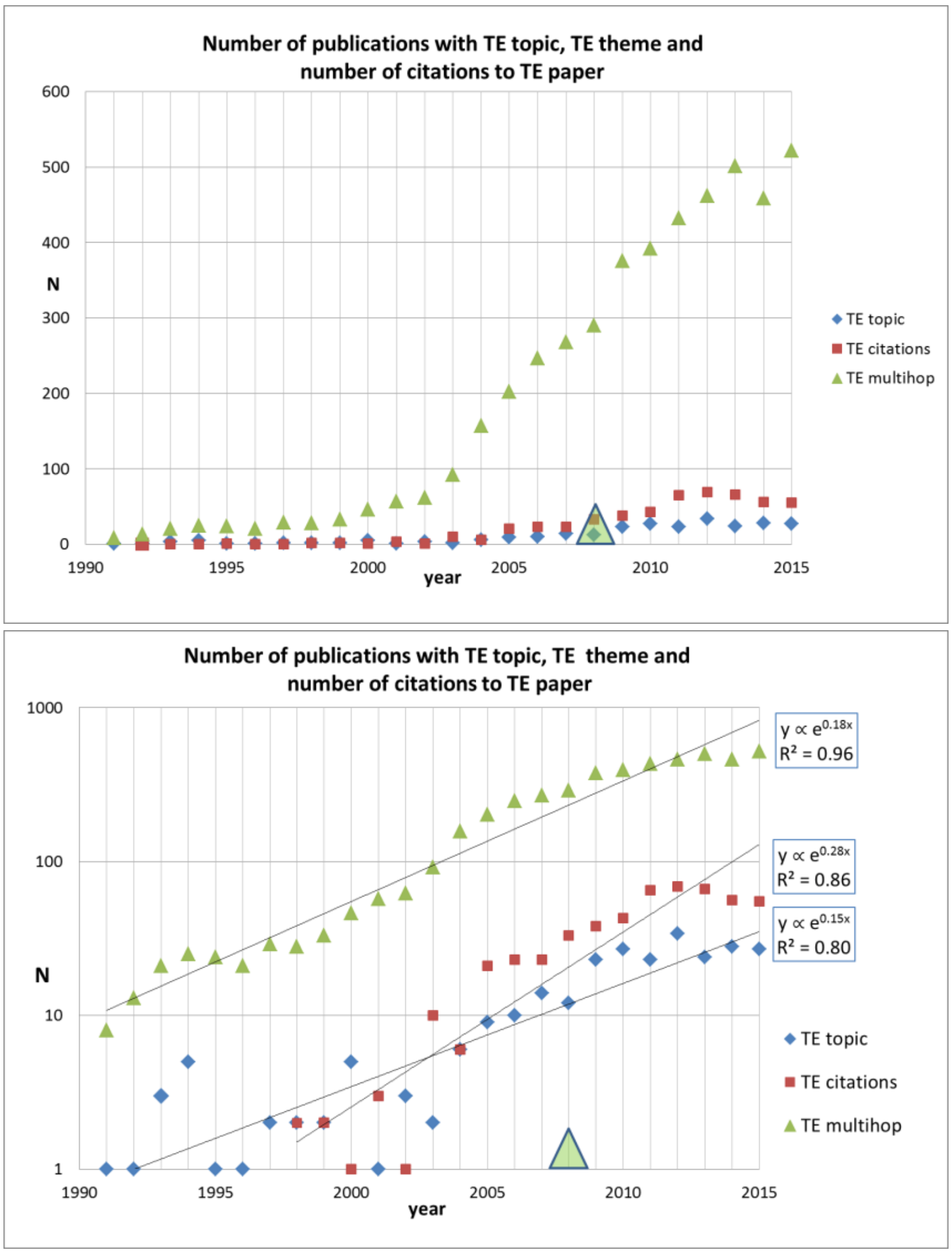

*Fig 10. Number of topic-related (TE topic) and theme-related papers (TE multihop) and number of citing papers. Upper part: linear representation; lower part: semi-log representation. Red squares: number of citations (citing papers); blue diamonds: number of topic-related publications; green triangles: number of theme-related papers. The large green triangle indicates the year of first citation in a patent.*

We present the time evolution of the topic-related, the theme-related papers as well as the citations to the TE SB-SNPR in Fig 10. Again we take 1991 as the starting year for our analysis. We observe a very similar, exponential growth in number of publications for the topic- and theme-related publications (exponents 0.15 and 0.18, respectively). However, the increase of papers citing the TE SB-SNPR is about twice as large (exponent 0.28).

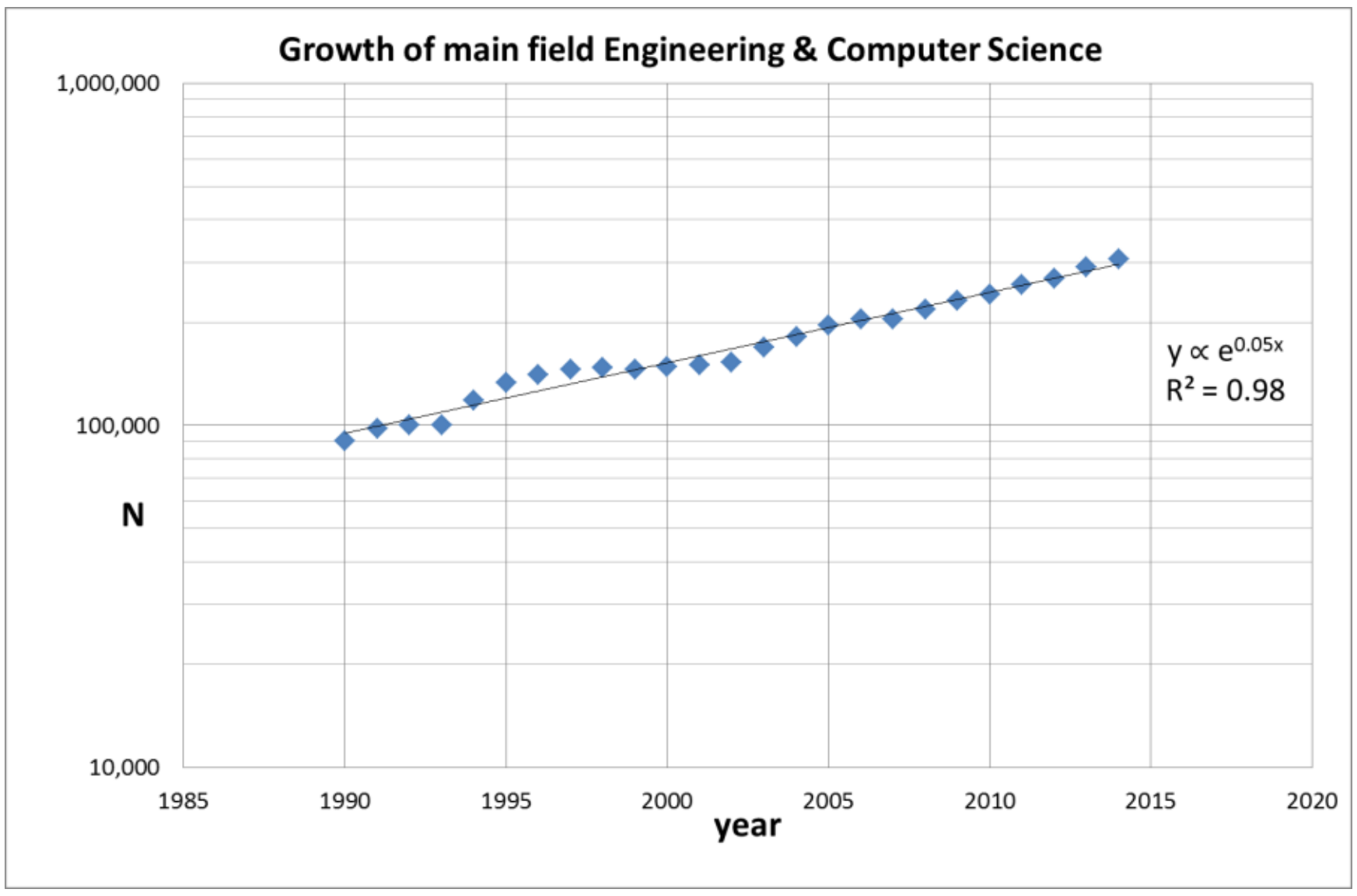

*Fig 11. Number of papers in Engineering and Computer Science, 1990-2015.*

Given the overall growth of the entire main field engineering & computer science with about 5 per cent per year in the period 1990-2015 as shown in Fig. 11, the growth of the TE SB-SNPR research theme is considerably faster indicating the importance of the theme.

## 4.4 Concept maps of topic-related and citing papers

Again following the analytical procedure developed in our foregoing paper (van Raan 2015) we applied the CWTS mapping tool VOS-viewer (for details see Van Eck and Waltman 2010) to construct concept maps of both the topic-related as well as the citing papers. It is however not very useful to create one map for the whole period after the awakening until now because this would lump together the concepts of publications over a period of about 15 years and make analysis of developments over time impossible. Therefore we will make a distinction between older and recent papers.

In Fig 12 we show in the upper part the concept map for the topic-related papers for the period until five years (up to 2006, in total 49 papers) after the awakening, and in the lower part the concept map for the recent five years (2011-2015, in total 136 papers).

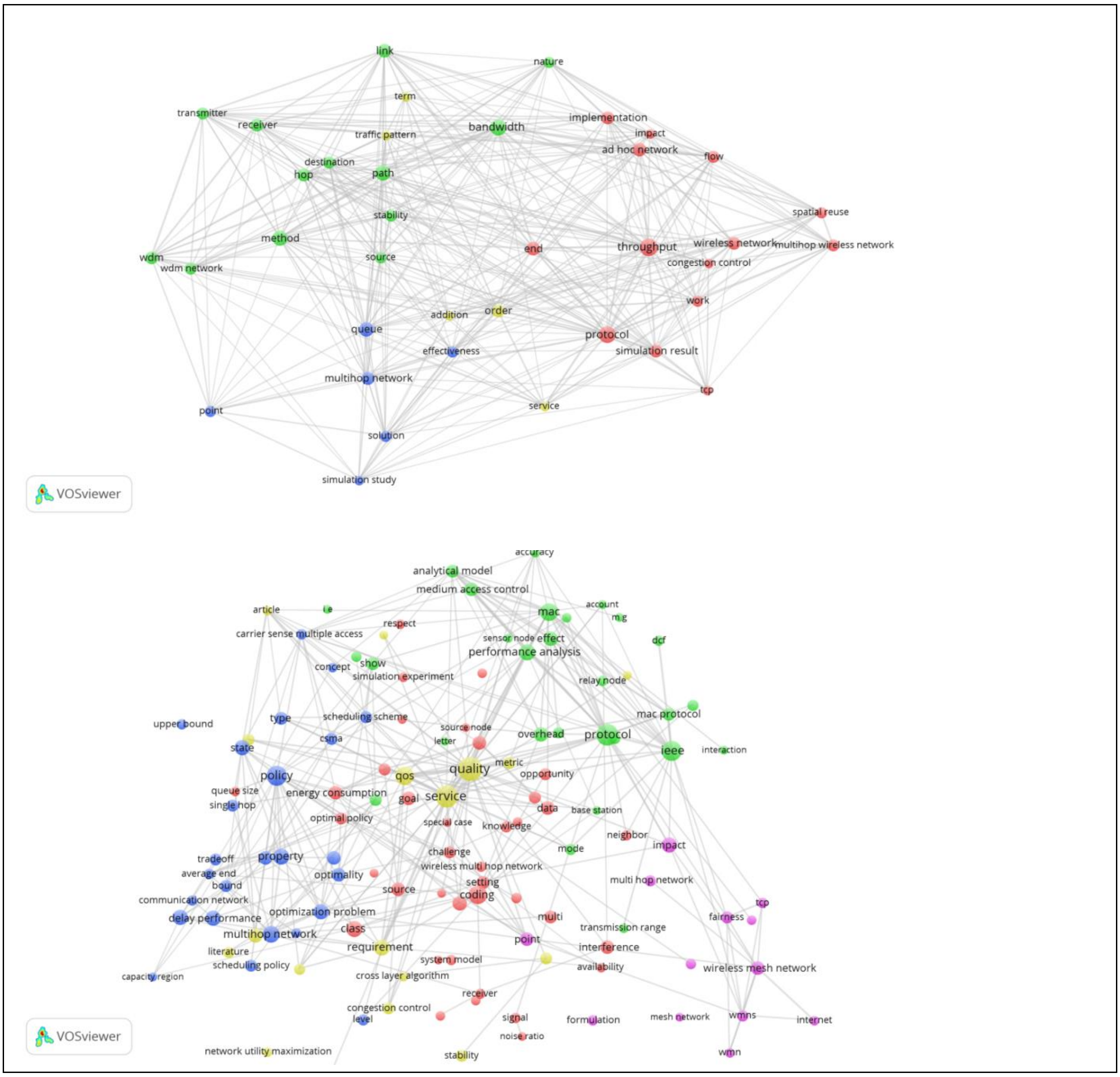

*Fig 12. Upper part: concept map of the topic-related papers in the period up to 2006; lower part: concept map of the topic-related papers in the period 2011-2015.*

We see that in the earlier period the focus is clearly on the topics as defined in the search string, namely queue and multihop networks, as well as on other important topics as throughput, bandwidth and wireless networks. In the recent period the map is more densely populated indicating the growth of the topic-related research activities. The defining topics remain but many other, particularly application-oriented topics have appeared on the scene such as quality, service, policy, protocol, coding, delay problems, access control.

In Fig 13 we show in the upper part the concept maps for the citing papers in the period 2002-2006 (in total 71 papers) and in the lower part the concept maps for the citing papers in the period 2011-2015 (in total 327 papers).

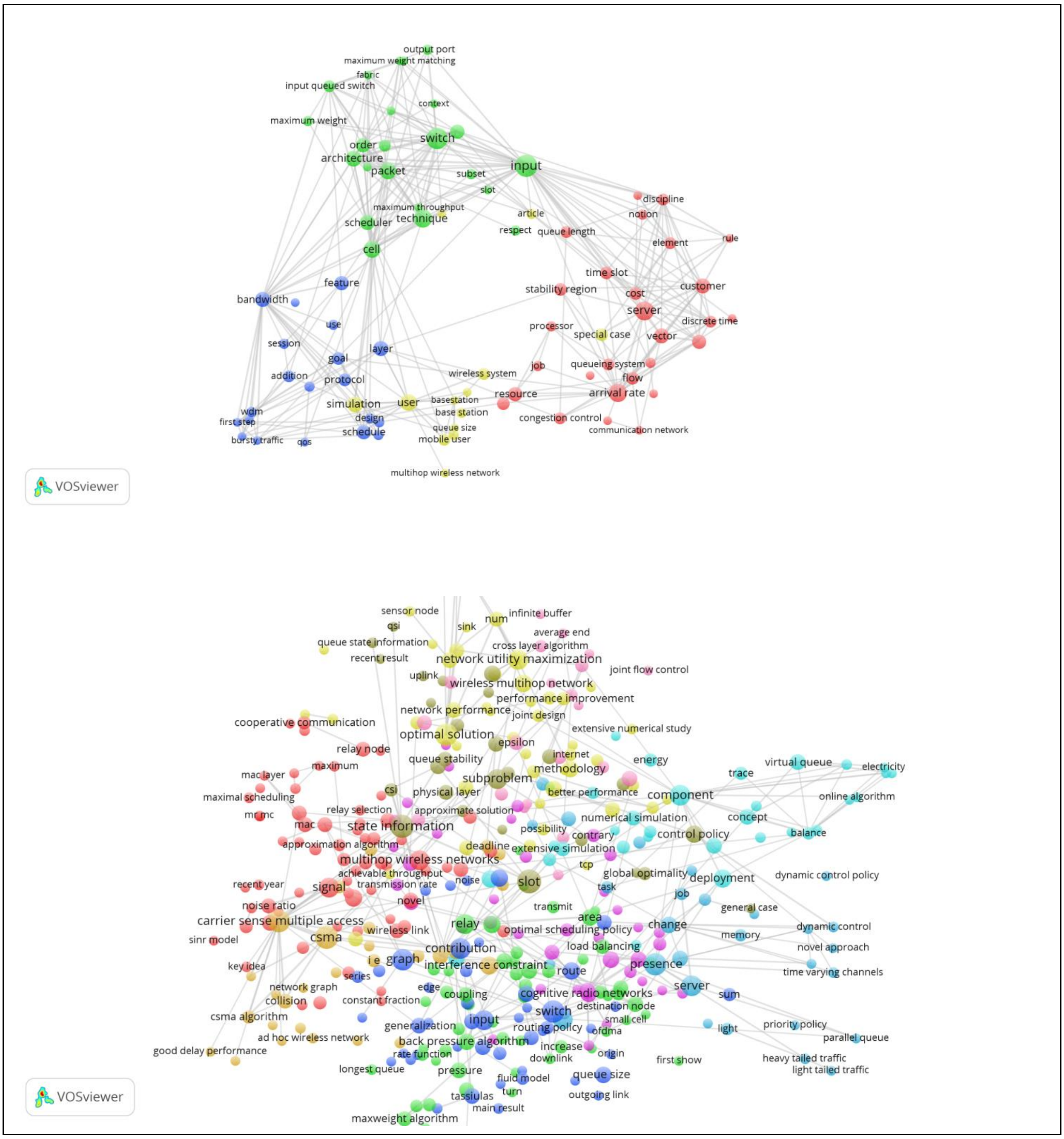

*Fig 13. Upper part: concept map of the citing papers in the period 2002-2006; lower part: concept map of the citing papers in the period 2011-2015.*

We see many similarities with the maps of the topic-related papers. For instance in the earlier period we find again queueing problems, multihop networks, throughput, bandwidth and wireless networks. But the scope is now broader and we observe for instance a cluster around customer-related problems. In the recent period the map of the citing papers is quite similar to the map of the topic-related problems but again the scope is broader which is well illustrated by a larger number of inks to many additional topics from central topics such as carrier sense multiple access (csma) and network performance and utility.

We observe the continuing dominating position of the most central concept, multihop wireless network. Clearly present is the core concept of the invention described in the TE

SB-SNPR paper: the back pressure algorithm. We quote Dr Neely[5]: "This paper [the TE SB-SNPR] models a multi-hop wireless network as a multi-queue system with transmission options described by a collection of link activation sets. It is the first to introduce Lyapunov drift for proving stability in a general multi-hop network. It introduces the important concepts of backpressure routing and maximum weight matching. This paper can be viewed as a foundation for scheduling for stability and maximum throughput in a data network." This description clearly shows the innovative role of the TE SB-SNPR and, given its technological importance, the work can be considered to be an invention.

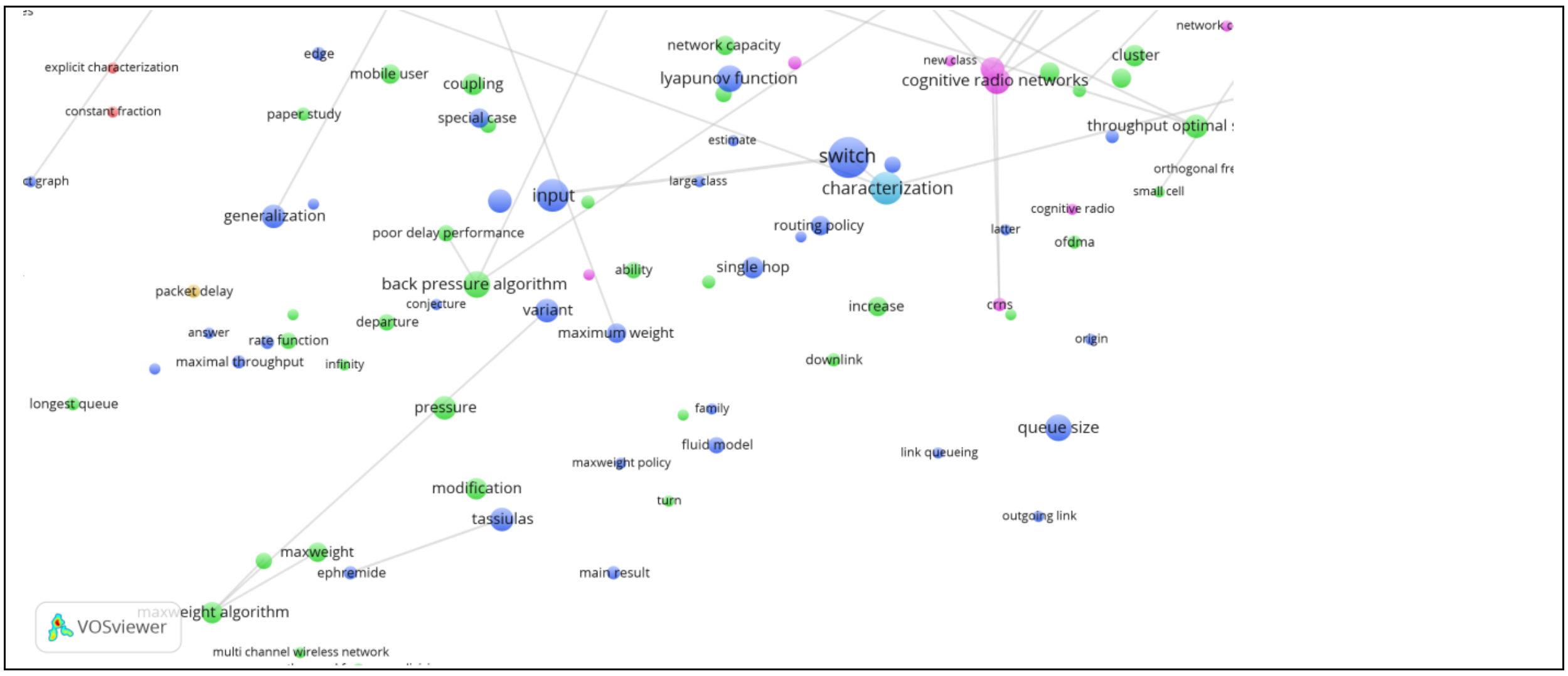

*Fig 14. Zoom into the backpressure algorithm region in the concept map of the citing papers in the period 2011-2015.*

When we zoom into the region around the concept backpressure algorithm in the lower part of Fig. 13, we observe most of the central issues mentioned in the above description. In the near vicinity of backpressure algorithm we find radio networks, max(imum) weight, routing, Lyapunov, mobile, and even the directly connected names of the authors Tassiulas and Ephremides, see Fig. 14.

## 4.5 Inventors as authors, authors as inventors

In Section 3.4 we discussed inventor-author self-citation of SB-SNPRs, i.e., inventors of a patented technology who are also author of the SB-SNPR cited in the patent. More general and broader types of the inventor-author link relate to (1) inventors who are also authors of papers, not necessarily cited in their own patents, and (2) the other way around, authors who are also inventors, and not necessarily cite their own papers in patents.

For type 1 the point of view is from the inventor. We use as an example the three patents in which the TE SB-SNPR is cited, see Table 7. We use similar headings in this table as in Table 5 and 6. We see that in total there are seven inventors. We find that these inventors are authors of 18 papers in the field of wireless networks in the period 2004-2014. The three most active inventor-authors are Radunivic (7 papers), Gkantsidis

[5] http://www-bcf.usc.edu/~mjneely/stochastic/index-old.html.

(4 papers, one together with Radunovic) and Marbach (4 papers). The three most cited of these 18 papers are all authored by Radunovic.

| Patent family | Patent publ | Patent title | Inventors | IPC codes | Appl year |
|---|---|---|---|---|---|
| 41608252 | US20080182493 | Path estimation in a wireless mesh network | Gkantsidis, Christos; Gunawardena, Dinan; Key, Peter B.; Radunovic, Bozidar | H04J3/14 | 2008 |
| 42106167 | US200913124019 | Delay and jitter limited wireless mesh network scheduling | Szymanski, Tadeusz H. | G01R31/08, H04L12/28, H04W72/12 | 2009 |
| 46316658 | US20100978151 | System and method for controlling data transmission in a multihop wireless network | Lotfinezhad, Mahdi; Marbach, Peter | H04L12/26 | 2010 |

*Table 7. Patents in which the BLSC SB-SNPR is cited.*

In Fig 15 the co-citation map of the 18 papers of the inventors is shown. The TE SB-SNPR is prominently present which means that it is also frequently cited in the papers of the inventor that cite the TE SB-SNPR in their patents.

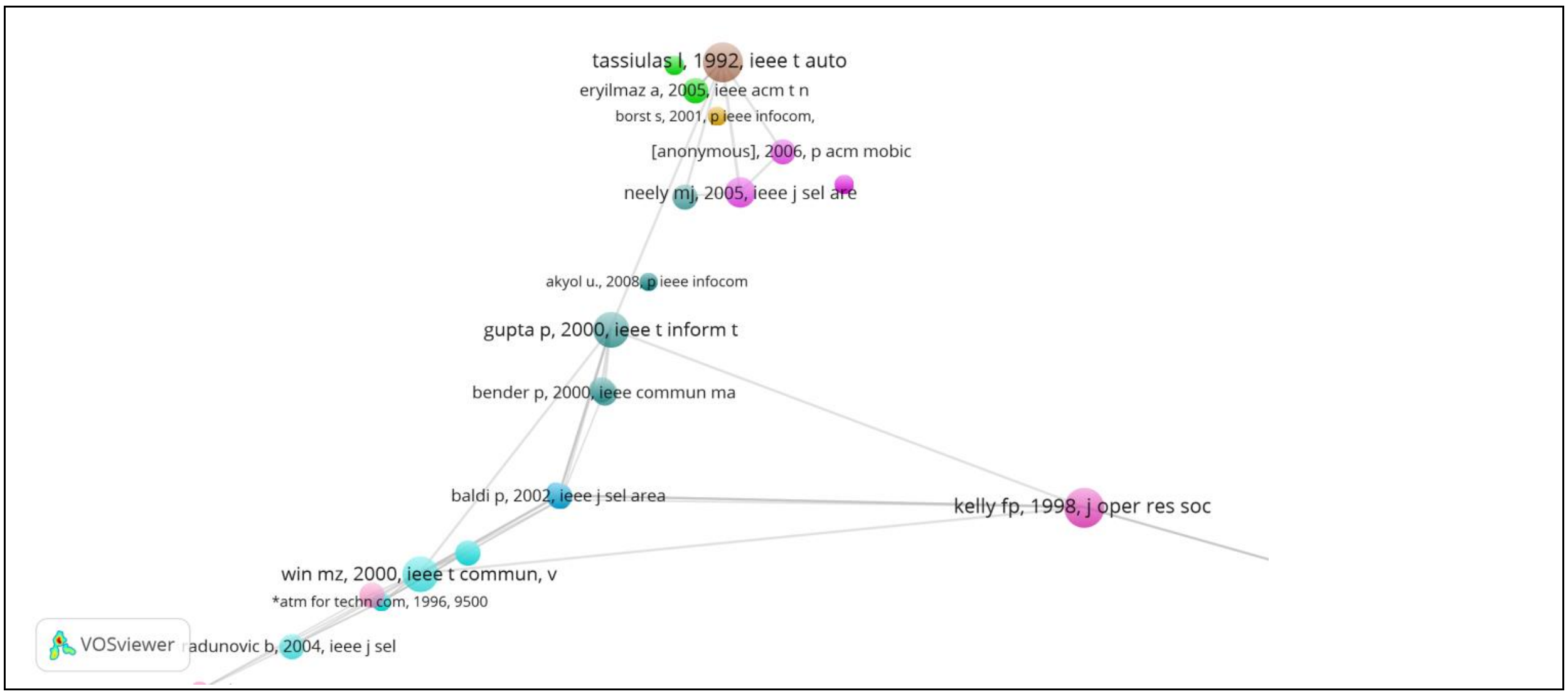

*Fig 15. Co-citation map of the papers of the inventors who cite the TE SB-SNPR in their patents.*

For type 2 the point of view is from the author. We take as an example the two authors of the TE SB-SNPR, Tassiulas and Ephremides. Both authors are inventors, we found for Tassiulas two patents granted by the United States Patent and Trademark Office (USPTO)[6], and for Ehremides eight patents granted by the USPTO[7].

From Table 7 it is clear that these inventor-authors do not cite in their own patents the TE SB-SNPR. But they may cite other papers of their own, for instance Tassiulas cites in US patent 7,961,702 Tassiulas and Sarkar (2002).

---

[6] Patent numbers US 6,377,812 (2002) and US 7,961,702 (2011).
[7] Patent numbers US 5,987,328 (1999), US 6,278,701 (2001), US 6,879,572 (2005), US 6,894,991 (2005), US 6,947,407 (2005), US 7,002,920 (2006), US 7,233,584 (2007), and US 8,542,579 (2013).

In forthcoming work we will investigate in more detail to what extent authors of Sleeping Beauties are inventors of technological developments described in later patents.

## 4.6 Identification of the prince

Earlier in this paper we suggested that two of the 25 first citing papers, Neely et al (2003) and Andrews et al (2003), are appropriate candidates for the role of prince. In order to make a reasonable decision we created a co-citation map of all citing papers of the TE SB-SNPR. The VoSViewer makes it possible to create maps with different thresholds thus enabling 'tunable co-citation analysis'. We use a high threshold to represent the strongest co-citation relations. This map is shown in Fig 16 and it clearly shows the central position of the TE SB-SNPR paper. Also here the different colors mark different research themes. We remind that in a co-citation map the references of the citing papers are clustered based on their co-occurrence in these citing papers. Because older citing papers can be references of recent citing papers, important older citing papers will show up in the co-citation map.

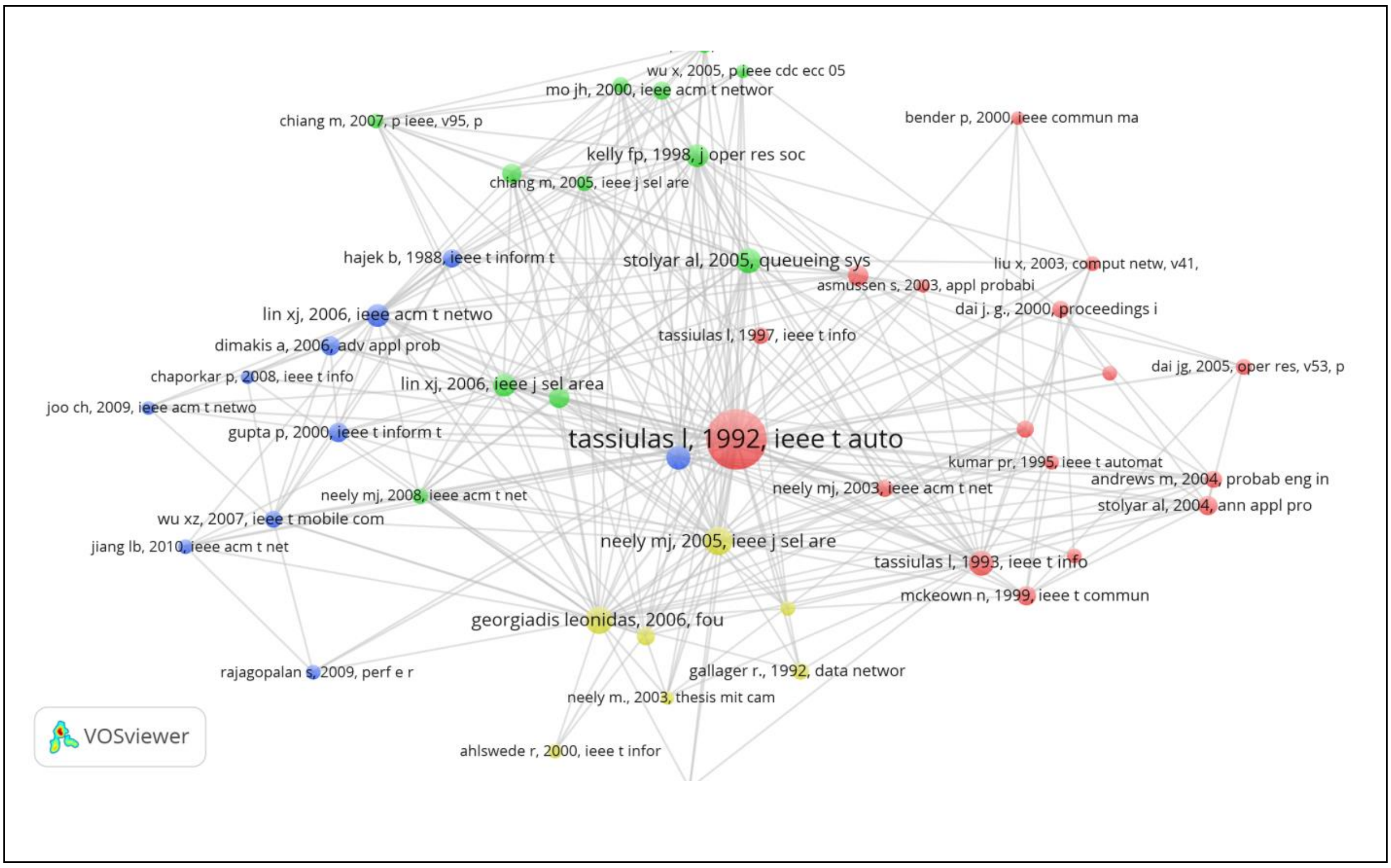

*Fig 16. Co-citation map of the papers citing the TE SB-SNPR.*

The size of the circles represents the occurrence of a specific reference within the total set of citing papers. As this total set of citing papers consists by definition of all papers citing the TE SB-SNPR, this paper has the largest circle size and there is no doubt that it takes a central position. Important papers frequently co-cited with the TE SB-SNPR are for instance the Neely et al (2005) paper and another paper by Tassiulas and Ephremides (1993). The Neely et al (2005) paper is an important paper on the backpressure algorithm developed in the TE SB-SNPR.

But the Neely et al (2005) paper is not the prince. In the figure we see that the earlier citing paper Neely et al (2003) is the oldest of all major (larger circles) papers close to the TE SB-SNPR. It is this earlier Neely et al paper we already considered a strong candidate for being the 'prince' on the basis of our observations of the first 25 papers citing the TE SB-SNPR.

We also constructed a co-citation map of the topic-related papers. As only 20% of these topic-related papers is also a paper citing the TE SB-SNPR, we can expect a considerable difference with the co-citation map of the citing papers. This is indeed clearly the case as shown in Fig 17. Although the TE SB-SNPR undoubtedly has a dominant position, the map is 'multipolar': particularly the highly cited Gupta et al (2000) paper on the capacity of wireless networks (cited 1,740 times, April 18, 2016) and the highly cited Bianchi et al (2000) paper on throughput analysis of wireless local area networks (cited 1,534 times, April 18, 2016) have central positions.

We find the Gupta paper also in the co-citation map of the papers citing the TE SB-SNPR (Fig 16, left side; the Bianchi paper appears in the map when tuning to a lower co-citation threshold). These papers however are less prominent in the co-citation map of the papers citing the SB-SNPR which indicates that they focus on related but still different themes (performance of wireless networks) of as compared to the TE paper (queueing and scheduling problems). In the topic-related papers co-citation map we again see the prince, the Neely et al paper (2003), and also here it is the only paper of the first 25 citing paper in the direct vicinity of the TE SB-SNPR.

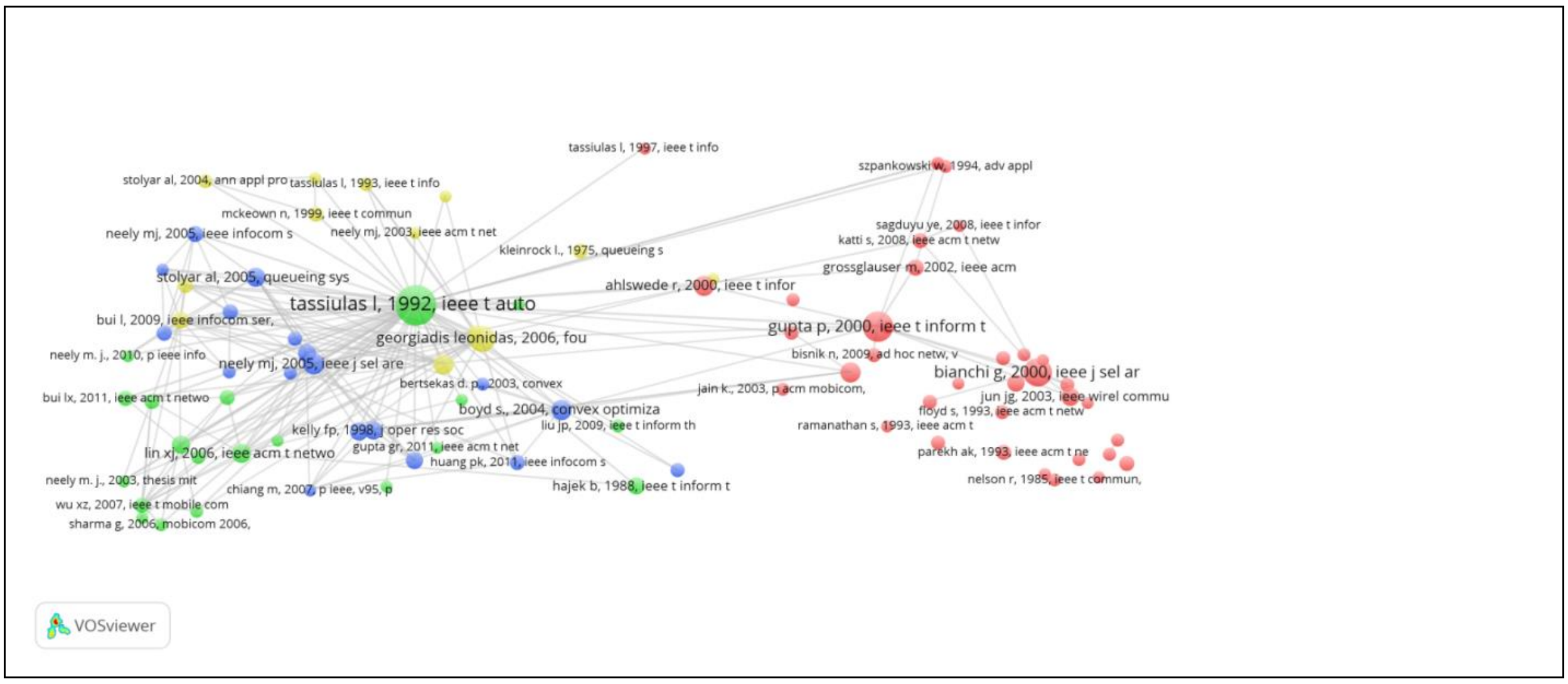

*Fig 17. Co-citation map of the topic-related papers.*

The above observations show the differences and similarities of the cognitive environment of a technologically relevant Sleeping Beauty (an SB-SNPR) in terms of its topic-related papers and its citing papers. We think that these 'two worlds' of relatedness (topic-related and citation-related) are a general phenomenon which is characteristic for any important scientific paper. As we have shown in this section, the analysis of this two worlds cognitive environment offers detailed information on the emerging impact of SB-SNPRs.

## 5. Concluding Remarks

In this paper we investigated characteristics of Sleeping Beauties that are cited in patents (SB-SNPRs). We found that patent citation may occur before or after the awakening and that the depth of the sleep, i.e., citation rate during the sleeping period, is no predictor for later scientific or technological impact of the Sleeping Beauty. A surprising finding is that Sleeping Beauties are significantly more cited in patents than 'normal' papers. In a comprehensive approach to explore the cognitive environment of a Sleeping Beauty we analyzed whether Sleeping Beauties cited by patents deal with new topics.

The awakening was studied in detail by analyzing the first group of papers that, after the sleeping period, cites the Sleeping Beauty. Concept maps and citation-based maps proved to be powerful tool to discover the prince and other important application-oriented work directly related to the Sleeping Beauty, for instance papers written by authors who cite Sleeping Beauties in both the patents of which they are the inventors, as well as in their scientific papers. These inventor-author relations in the context of Sleeping Beauties may provide more insight into the phenomenon of Sleeping Inventions and we are currently focusing on this theme.

The average time lag between the publication year of an SB-SNPR and its first citation in a patent appears to decrease in the 1980's and early 1990's. This would mean that SBs with technological importance, perhaps potential inventions, are 'discovered' increasingly earlier. This also could imply a decreasing probability that a Sleeping Beauty is cited by a patent because the technological relevance of a publication is recognized before it can become a Sleeping Beauty. Our forthcoming work on more recent, shorter Sleeping Beauties may shed more light on this issue.

**Acknowledgements**

The author thanks his colleagues Nees-Jan van Eck for developing and writing the Sleeping Beauties algorithm and Jos Winnink for the collection and first analysis of the patent data.

## Appendix

Table A1: Definition of the main fields physics (upper part), chemistry (middle part), and engineering & computer science (lower part) based on WoS journal categories.

| | |
|---|---|
| **Physics** | |
| 1 | ACOUSTICS |
| 20 | ASTRONOMY & ASTROPHYSICS |
| 27 | BIOPHYSICS |
| 35 | THERMODYNAMICS |
| 152 | MATERIALS SCIENCE, BIOMATERIALS |
| 153 | MATERIALS SCIENCE, CHARACTERIZATION & TESTING |
| 154 | MATERIALS SCIENCE, COATINGS & FILMS |
| 155 | MATERIALS SCIENCE, COMPOSITES |
| 156 | MATERIALS SCIENCE, TEXTILES |
| 159 | METEOROLOGY & ATMOSPHERIC SCIENCES |
| 168 | NUCLEAR SCIENCE & TECHNOLOGY |
| 175 | OPTICS |
| 185 | PHYSICS, APPLIED |
| 187 | PHYSICS, FLUIDS & PLASMAS |
| 188 | PHYSICS, ATOMIC, MOLECULAR & CHEMICAL |
| 189 | PHYSICS, MULTIDISCIPLINARY |
| 190 | PHYSICS, CONDENSED MATTER |
| 192 | PHYSICS, NUCLEAR |
| 193 | PHYSICS, PARTICLES & FIELDS |
| 195 | PHYSICS, MATHEMATICAL |
| **Chemistry** | |
| 23 | BIOCHEMICAL RESEARCH METHODS |
| 24 | BIOCHEMISTRY & MOLECULAR BIOLOGY |
| 36 | CHEMISTRY, APPLIED |
| 37 | CHEMISTRY, MEDICINAL |
| 38 | CHEMISTRY, MULTIDISCIPLINARY |
| 39 | CHEMISTRY, ANALYTICAL |
| 40 | CHEMISTRY, INORGANIC & NUCLEAR |
| 41 | CHEMISTRY, ORGANIC |
| 42 | CHEMISTRY, PHYSICAL |
| 57 | CRYSTALLOGRAPHY |
| 63 | GEOCHEMISTRY & GEOPHYSICS |
| 71 | ELECTROCHEMISTRY |
| 198 | POLYMER SCIENCE |
| **Engineering & Computer Science** | |
| 6 | ENGINEERING, AEROSPACE |
| 28 | BIOTECHNOLOGY & APPLIED MICROBIOLOGY |
| 44 | COMPUTER SCIENCE, ARTIFICIAL INTELLIGENCE |
| 46 | COMPUTER SCIENCE, CYBERNETICS |
| 47 | COMPUTER SCIENCE, HARDWARE & ARCHITECTURE |

| | | |
|---|---|---|
| | 48 | COMPUTER SCIENCE, INFORMATION SYSTEMS |
| | 49 | COMMUNICATION |
| | 50 | COMPUTER SCIENCE, INTERDISC APPLICATIONS |
| | 51 | COMPUTER SCIENCE, SOFTWARE ENGINEERING |
| | 52 | COMPUTER SCIENCE, THEORY & METHODS |
| | 54 | CONSTRUCTION & BUILDING TECHNOLOGY |
| | 75 | ENERGY & FUELS |
| | 76 | ENGINEERING, MULTIDISCIPLINARY |
| | 77 | ENGINEERING, BIOMEDICAL |
| | 78 | ENGINEERING, ENVIRONMENTAL |
| | 79 | ENGINEERING, CHEMICAL |
| | 80 | ENGINEERING, INDUSTRIAL |
| | 81 | ENGINEERING, MANUFACTURING |
| | 82 | ENGINEERING, MARINE |
| | 83 | ENGINEERING, CIVIL |
| | 84 | ENGINEERING, OCEAN |
| | 85 | ENGINEERING, PETROLEUM |
| | 86 | ENGINEERING, ELECTRICAL & ELECTRONIC |
| | 87 | ENGINEERING, MECHANICAL |
| | 97 | FOOD SCIENCE & TECHNOLOGY |
| | 119 | INSTRUMENTS & INSTRUMENTATION |
| | 131 | OPERATIONS RESEARCH & MANAGEMENT SCIENCE |
| | 145 | MEDICAL LABORATORY TECHNOLOGY |
| | 147 | METALLURGY & METALLURGICAL ENGINEERING |
| | 168 | NUCLEAR SCIENCE & TECHNOLOGY |
| | 173 | REMOTE SENSING |
| | 186 | IMAGING SCIENCE & PHOTOGRAPHIC TECHNOLOGY |
| | 222 | TELECOMMUNICATIONS |
| | 227 | TRANSPORTATION |
| | 237 | MINING & MINERAL PROCESSING |
| | 242 | TRANSPORTATION SCIENCE & TECHNOLOGY |
| | 244 | AGRICULTURAL ENGINEERING |
| | 245 | CRITICAL CARE MEDICINE |
| | 247 | ENGINEERING, GEOLOGICAL |
| | 248 | INTEGRATIVE & COMPLEMENTARY MEDICINE |
| | 251 | ROBOTICS |
| | 252 | NANOSCIENCE & NANOTECHNOLOGY |
| | 257 | CELL & TISSUE ENGINEERING |